\documentclass[journal]{IEEEtran}
\usepackage{graphicx}
\usepackage{subfigure}
\usepackage{amsmath}
\usepackage{multirow}
\usepackage{float}
\usepackage{colortbl}
\usepackage{xcolor}
\usepackage{multirow}
\usepackage{booktabs}
\usepackage{amsfonts,amssymb}
\usepackage{hyperref}
\usepackage{fontawesome}
\hypersetup{
	colorlinks=true,
	linkcolor=blue,
	urlcolor=blue,
	citecolor=blue,
}

\begin{document}
\title{Prototype-Based Information Compensation Network for Multi-Source Remote Sensing Data Classification}

\author{
  Feng Gao, \emph{Member}, \emph{IEEE},
  Sheng Liu, 
  Chuanzheng Gong,
  Xiaowei Zhou, \emph{Member}, \emph{IEEE},
  Jiayi Wang, \\
  Junyu Dong, \emph{Member}, \emph{IEEE},
  Qian Du, \emph{Fellow}, \emph{IEEE}
\thanks{This work was supported in part by the National Science and Technology Major Project of China under Grant 2022ZD0117202, in part by the National Natural Science Foundation of Shandong province under Grant ZR2024MF020, in part by Natural Science Foundation of Qingdao under Grant 23-2-1-222-zyyd-jch, and in part by the Postdoctoral Fellowship Program of CPSF under Grant GZC20241614.
\emph{(Corresponding author: Xiaowei Zhou)}} %
\thanks{F. Gao, S. Liu, C. Gong, X. Zhou, J. Wang and J. Dong are with the State Key Laboratory of Physical Oceanography, Ocean University of China, Qingdao 266100, China. 

Qian Du is with the Department of Electrical and Computer Engineering, Mississippi State University, Starkville, MS 39762 USA.}}

\markboth{IEEE TRANSACTIONS ON GEOSCIENCE AND REMOTE SENSING}
{Shell}

\maketitle

\begin{abstract}

Multi-source remote sensing data joint classification aims to provide accuracy and reliability of land cover classification by leveraging the complementary information from multiple data sources. Existing methods confront two challenges: inter-frequency multi-source feature coupling and inconsistency of complementary information exploration. To solve these issues, we present a Prototype-based Information Compensation Network (PICNet) for land cover classification based on HSI and SAR/LiDAR data. Specifically, we first design a frequency interaction module to enhance the inter-frequency coupling in multi-source feature extraction. The multi-source features are first decoupled into high- and low-frequency components. Then, these features are recoupled to achieve efficient inter-frequency communication.  Afterward, we design a prototype-based information compensation module to model the global multi-source complementary information. Two sets of learnable modality prototypes are introduced to represent the global modality information of multi-source data. Subsequently, cross-modal feature integration and alignment are achieved through cross-attention computation between the modality-specific prototype vectors and the raw feature representations. Extensive experiments on three public datasets demonstrate the significant superiority of our PICNet over state-of-the-art methods. The codes are available at \url{https://github.com/oucailab/PICNet}.

\end{abstract}


\IEEEpeerreviewmaketitle

\section{Introduction}

\IEEEPARstart{I}{n} recent decades, multi-source remote sensing sensors have achieved remarkable progress, exerting a substantial influence on crop mapping \cite{aj25jstars} \cite{qjh24tgrs}, environmental monitoring \cite{ln25tgrs} \cite{zgl24grsl}, and urban planning \cite{ja25jstars} \cite{chh24tgrs}. These sensors have provided a huge amount of multi-source remote sensing images and play a crucial role in modern Earth observation \cite{fyn24tgrs} \cite{wsh25tgrs}.

One of the key advantages of multi-source data is their ability to provide complementary information. For example, hyperspectral image (HSI) provides spectral information \cite{wjj21tgrs} \cite{pzj25tgrs}, enabling the identification of land cover types and vegetation health. Synthetic aperture radar (SAR) data can penetrate through clouds, allowing for the observation of the Earth's surface even in adverse weather conditions and providing information about soil moisture and ocean surface roughness \cite{qxf22grsl}. Furthermore, LiDAR data provides high spatial and vertical resolution for generating accurate elevation models and is a valuable tool for 3D mapping and forestry applications \cite{ywy24tgrs}. The complementary information provided by multi-source images enables a wide range of tasks and applications, such as land cover classification, disaster monitoring, and urban mapping \cite{yb24jstars}. Among these applications, land cover classification is a cost-effective alternative to extensive ground-based investigations, and it is essential for understanding the Earth's surface effectively. Hence, in this paper, we focus on the land cover classification task using multi-source remote sensing images, specifically HSI and LiDAR/SAR data.

\begin{figure}[]
\centering
\includegraphics[width=0.8\linewidth]{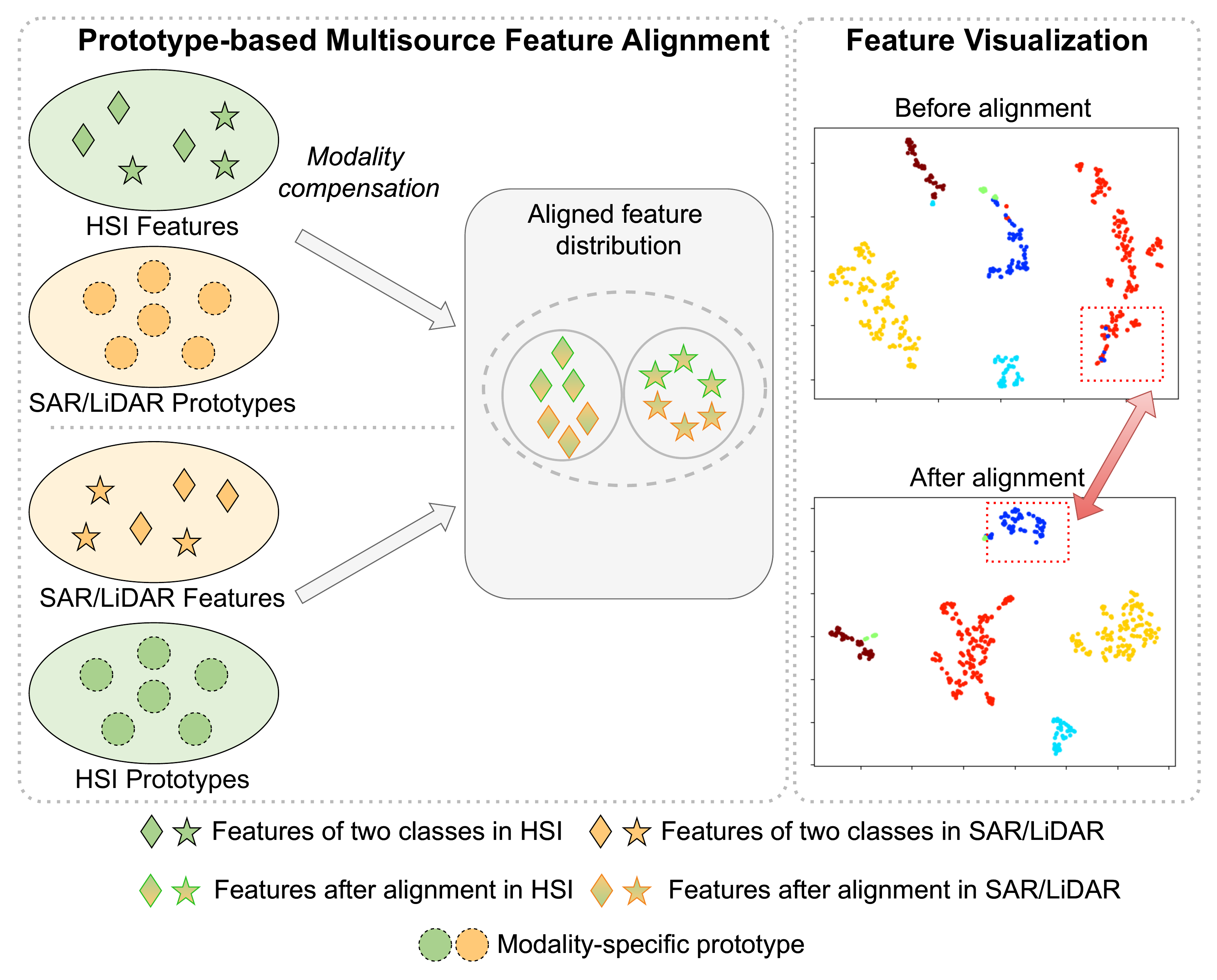} \caption{The motivation of the proposed Prototype-based Information Compensation Network (PICNet). Two sets of learnable modality-specific prototypes are introduced to store the global modality features, which can be used to compensate for the multi-source feature alignment.}
\label{fig_motivation}
\end{figure}

Numerous deep learning-based methods have been designed to exploit the complementary information contained in the multi-source remote sensing images \cite{gyh24tip}. Coupled CNNs \cite{hrl20tgrs}, fractional Gabor CNNs \cite{zxd22tgrs}, nearest neighbor contrastive learning \cite{wm23nncnet}, and cross-attention fusion \cite{gyh23tgrs} have been introduced for mutli-source data classification. To capture long-range dependencies and global contextual information, Transformer has been applied for multi-source data classification recently. Hierarchical Transformer \cite{xzx22tip}, local interactive network \cite{zyw23jstars}, attention-free network \cite{sl24tgrs}, and multi-head selection network \cite{nk24jstars} are proposed for multi-source data joint classification. 

\begin{figure}[]
\centering
\includegraphics[width=0.8\linewidth]{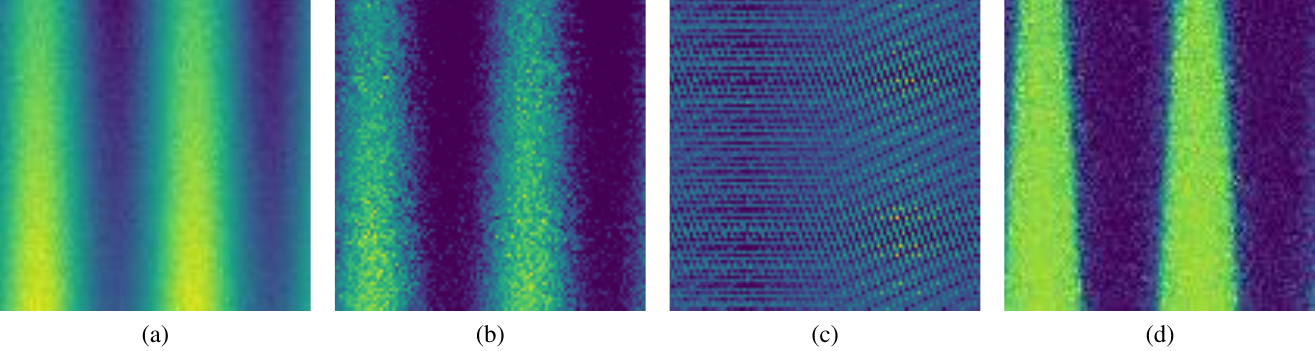} \caption{Comparative analysis of the efficacy between the Frequency Interaction Module (FIM) proposed in this study and conventional inter-frequency coupling methodologies. (a) Original input image, (b) Feature map before frequency interaction, (c) Feature representation after Wavelet transformation, (d) Enhanced feature map subsequent to FIM.}
\label{fig_fre_compare}
\end{figure}

Despite the remarkable performance achieved by previous studies, we argue that building an effective multi-source data classification model remains a non-trivial task due to the following two challenges: 

\emph{\textbf{(1) Inconsistency in complementary information exploration during multi-source feature fusion.}} In previous methods, cross-attention\cite{gyh23tgrs} or bilinear attention\cite{wm23nncnet} mechanisms are applied based solely on the samples of the current mini-batch, failing to exploit the full spectrum of complementary information available across different mini-batches. Therefore, it is imperative to model and integrate global multi-source complementary information to facilitate more effective alignment between HSI and SAR/LiDAR data.

\emph{\textbf{(2) Limited inter-frequency coupling in multi-source feature extraction.}} Existing frequency-based methods directly employ a global frequency filter for HSI and SAR/LiDAR feature extraction\cite{zz24icassp} \cite{nk24tgrs}, which overlooks the rich, complementary information embedded in distinct frequency components. As shown in Fig. \ref{fig_fre_compare}, although the traditional wavelet transform can capture some frequency information, it leads to information loss and does not effectively capture the complementary information between different frequency components. Consequently, enhancing the inter-frequency recoupling is crucial for capturing these nuances and achieving robust cross-modal feature extraction.

To address the above-mentioned two challenges, we propose a \textbf{P}rototype-based \textbf{I}nformation \textbf{C}ompensation \textbf{Net}work (PICNet) for land cover classification based on HSI and SAR/LiDAR data. The overall framework of the proposed PICNet is illustrated in Fig. \ref{fig_frame}. Specifically, we design a Prototype-based Information Compensation Module (PICM) to model the global multi-source complementary information. As shown in Fig. \ref{fig_motivation}, two sets of learnable modality prototypes are introduced to represent the global modality information of HSI and SAR/LiDAR data. Subsequently, cross-modal feature integration and alignment are achieved through cross-attention computations between the modality-specific prototype vectors and the raw feature representations. This global sample feature compensation mechanism effectively harmonizes the exploration of complementary information across multi-source feature fusion, ensuring consistent feature integration. This global sample feature compensation mechanism is a novel contribution. Unlike previous methods that rely solely on cross-attention or bilinear attention mechanisms based on the samples of the current mini-batch, our PICM leverages global multi-source complementary information, ensuring consistent feature integration across different mini-batches. This innovation is crucial for achieving robust and reliable multi-source feature fusion.

Afterward, we design a Frequency Interaction Module (FIM) to enhance the inter-frequency coupling in multi-source feature extraction. The multi-source features are first decoupled into high- and low-frequency components. Then, these features are recoupled to achieve efficient inter-frequency communication. This aggregation strategy, which integrates cross-modal and multi-frequency information.This approach is distinct from traditional frequency-based methods that directly employ global frequency filters for feature extraction, which often overlook the rich, complementary information embedded in distinct frequency components. By recoupling the high- and low-frequency components, as shown in Fig. \ref{fig_fre_compare}, our method ensures seamless inter-frequency information communication, thereby addressing the issue of limited inter-frequency coupling in existing methods. To verify the effectiveness of the proposed PICNet, we conducted extensive experiments on the Augsburg \cite{essd-15-113-2023}, Berlin \cite{okujeni2016berlin}, and Houston 2018 \cite{8328995} datasets. Experimental results demonstrate that the proposed PICNet can yield better classification performance as compared to the state-of-the-art methods.

To the best of our knowledge, it is the first work on exploring global multi-source complementary information via modality-specific prototypes. The main contributions of this paper are threefold:

\begin{itemize} 

\item \textbf{Multi-source feature alignment contribution.} We propose PICM to model the global multi-source complementary information via two sets of  modality-specific prototype vectors. This approach effectively addresses the challenge of inconsistent complementary information exploration during multi-source feature fusion. By leveraging cross-attention computation between the prototype vectors and the raw feature representations, PICM ensures consistent feature integration across different mini-batches, leading to more robust and reliable multi-source feature alignment.

\item \textbf{Feature representation contribution.} We design FIM to facilitate the inter-frequency coupling in multi-source feature extraction. This module decouples the multi-source features into high- and low-frequency components and then recouples them to achieve seamless inter-frequency information communication. This strategy overcomes the limitations of traditional frequency-based methods.

\item \textbf{Experimental contribution.} Our PICNet demonstrates superior classification performance compared to the state-of-the-art methods on three public multi-source remote sensing classification datasets. The codes and settings will be made publicly available to benefit other researchers.

\end{itemize}

\section{Related Works}

\subsection{Multi-Source Remote Sensing Image Classification}

Single-source remote sensing image classification is highly susceptible to fluctuations in image quality, as the acquisition process can be influenced by factors such as weather conditions, lighting variations, and inherent sensor limitations. Multi-source remote sensing image classification addresses these challenges by integrating images from multiple sensors, thereby reducing the impact of individual sensor limitations and enhancing model robustness \cite{pedram15fuse}. In recent years, numerous deep learning-based approaches have been developed to advance the classification of multi-source remote sensing data, leveraging the complementary strengths of different data sources \cite{10271328}\cite{10050424}.

Cai et al. \cite{Cai2024ESA} introduced a graph attention-based fusion module and constructed a multi-source feature map to address the long-range dependency in HSI. Liu et al. \cite{Liu2021KBS} designed a spectral and spatial attention feature fusion module specifically designed for HSI and LiDAR data, achieving significant improvements in fusion performance. Hu et al. \cite{Hu2024JSTARS} developed a decision-level fusion method based on Dempster-Shaffer theory to improve classification accuracy. Afterward, Ni et al. \cite{nk25grsl} used selective convolution kernels and spectral-spatial interactive Transformer for multi-source data classification. Qu et al. \cite{qjh24tgrs} presented a shared-private decoupling-based multilevel feature alignment method for HSI and LiDAR data classification, which uses a graph Transformer-based class-balanced pseudo-label generation strategy for iterative model training. Zhang et al. \cite{zgl24grsl} employed the state-space space (SSM) model to dynamically update multi-modal state information to fuse HSI and LiDAR data. Gao et al. \cite{gf25tgrs} proposed the multi-scale feature fusion Mamba for HSI and SAR joint classification. It employed the multi-scale strategy to reduce computational cost and alleviate feature redundancy in multiple scanning routes, ensuring efficient spatial feature modeling. Hong et al. \cite{HONG2023113856} proposed the HighDAN network. This network aims to enhance the generalization ability of AI models in cross-city remote sensing applications. HighDAN retains the spatial topological structure of urban scenes through a high-to-low resolution fusion method and uses adversarial learning to bridge the gap caused by differences in remote sensing image representations across cities. Additionally, it employs Dice loss to mitigate class imbalance issues arising from cross-city factors. Xi et al. \cite{10684809} proposed the CTF-SSCL network, which combines CNN and Transformer with semisupervised contrastive learning to enhance few-shot hyperspectral image classification. This network uses a lightweight spatial-spectral interactive convolution module and a multi-scale transformer to extract local and global features, and employs a semisupervised contrastive loss to optimize model performance. In addition, hierarchical Transformer \cite{xzx22tip}, local interactive network \cite{zyw23jstars}, attention-free network \cite{sl24tgrs}, multi-head selection network \cite{nk24jstars}, hashing-based metric learning \cite{sww23tgrs}, and masked auto-encoder \cite{ljy23tgrs} have been proposed for multi-source data joint classification. 

Despite these advancements, existing methods primarily focus on learning shared features of current mini-batch samples during training. The multi-source complementary information from different mini-batches is not fully exploited. Our PICNet captures global multi-source complementary information via modality-specific prototype vectors. Cross-modal feature alignment is achieved
from the global view, which is meaningful for effective and reliable multi-source feature fusion.

\subsection{Frequency Analysis in Remote Sensing Image Classification}

Frequency analysis is a widely used method for feature extraction in remote sensing image classification \cite{lyq25tgrs}. By separately analyzing high-frequency and low-frequency components, it enables the extraction of detailed and structural information to enhance classification performance.

Bai et al. \cite{Bai2022TGRS} leveraged OctConv to extract fine-grained frequency features, significantly improving CNN performance by integrating multi-frequency information. Dong et al. \cite{Dong2024TGRS} introduced a low-frequency reconstruction target that directs the model's attention toward fundamental ground object features while minimizing the influence of extraneous details in remote sensing images.  Similarly, Su et al. \cite{Su2021JSTARS} applied a two-dimensional discrete cosine transform to the original feature map, optimizing the extraction of effective eigenvalues across channels and enabling robust processing of noisy remote sensing images without additional filtering. To address challenges of high inter-class similarity and intra-class diversity, Zhang et al. \cite{Zhang2024TMM} proposed a customized multimodal fine-tuning strategy based on frequency distribution. This method integrates local-global frequency information, improving the recognition of remote sensing scenes. Li et al. \cite{fd2net} proposed FD2-Net to capture the unique frequency representations of complementary information across multi-modal visual spaces. Wang et al. \cite{10630569} proposed FDNet, a pure frequency domain deep learning model for remote sensing scene classification. This model integrates wavelet transform into the downsampling process, effectively preserving texture information that is often lost in traditional spatial domain downsampling methods.

\begin{figure*}[t]
\begin{center}
\includegraphics[width=0.8\linewidth]{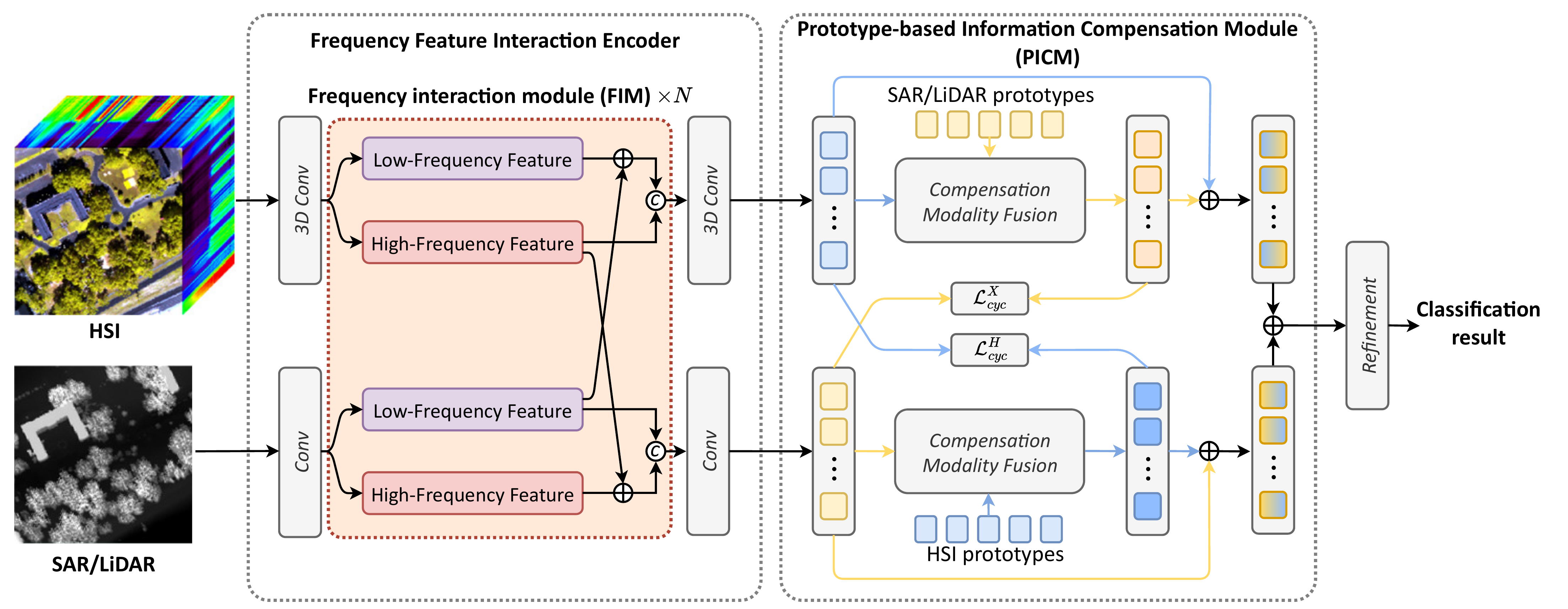} 
\end{center}
\caption{The overall framework of the proposed PICNet. It consists of two key components: (1) The frequency feature interaction encoder which leverages the Frequency Interaction Module (FIM) to decouple the frequency information of multi-source data and use the dominant features of one modality to enhance the complementary feature of the other. (2) The Prototype-based Information Compensation Module (PICM) is responsible for compensating the missing modality-specific information, which is achieved by cross-attention computation between the prototype vectors and the raw feature representations.}
\label{fig_frame}
\end{figure*}

The Laplacian pyramid decomposes the original image into multiple scaled layers through a series of downsampling and upsampling operations, with each layer representing distinct frequency ranges. Inspired by this, Zhang et al. \cite{Zhang2022TGRS} developed a CNN-based Laplacian high-frequency convolution block that employs a trainable Laplacian operator to extract valuable high-frequency features.  Liu et al. \cite{Liu2020TNNLS} introduced the C-CNN architecture, which separates input images into low-pass and high-pass subbands using the Laplacian pyramid. The network further employs a directional two-dimensional filter bank to reconstruct the original signal with minimal sample representation.

Both Fourier and wavelet transform provide effective mechanisms for transforming remote sensing images from the spatial domain to the frequency domain \cite{Yang2023TGRS, Morales2017IGARSS, Yang2024TGRS}, revealing the intensity and distribution of frequency components for targeted analysis. Morales et al. \cite{Morales2017IGARSS} proposed a method to approximate kernel functions in Gaussian process classification, integrating standard random Fourier feature approximation to enhance classification efficiency, particularly for large-scale remote sensing data. Yang et al. \cite{Yang2023TGRS} developed an adaptive discrete wavelet transform to extract global frequency multi-scale texture features, which significantly improved scene classification in areas with substantial size variation. Expanding on this, Yang et al. \cite{Yang2024TGRS} designed an adaptive multi-frequency graph feature learning module that combines graph convolution and graph wavelet convolution techniques to capture multi-scale low-frequency and high-frequency features from HSI and LiDAR data.

While many existing approaches rely on wavelet or Fourier transforms for feature extraction, these transform-based methods often lead to information loss during feature learning. In this paper, we employ a pooling-based frequency feature separation for multi-source feature extraction, which is rather computationally efficient. In addition, the high- and low-frequency components from multi-source data are recoupled multiple times for feature enhancement.

\section{Methodology}

\subsection{Overall Network Architecture}

In this subsection, we introduce the proposed Prototype-based Information Compensation Network (PICNet) in detail. As depicted in Fig. \ref{fig_frame}, PICNet consists of two key components: (1) The frequency feature interaction encoder which leverages the Frequency Interaction Module (FIM) to decouple the frequency information of multi-source data and use the dominant features of one modality to enhance the complementary feature of the other. (2) The Prototype-based Information Compensation Module (PICM) is responsible for compensating the missing modality-specific information, which is achieved by cross-attention computation between the prototype vectors and the raw feature representations.

The features generated by the prototype-based information compensation module are fed into MLP to compute the final classification results.  The modality consistency loss and cross-entropy loss guide the network training process. These loss functions enforce alignment between multi-source features and improve the scalability of the network.

\subsection{Frequency Feature Interaction Encoder}

Formally, let $\mathbf{I}^H$ denote the input HSI, and $\mathbf{I}^X$ denote the input LiDAR/SAR data. Initially, 3D convolution is employed for $\mathbf{I}^H$ preprocessing, and 2D convolution is used for $\mathbf{I}^X$ preprocessing. Next, the multi-source features are fed into the Frequency Interaction Module (FIM) for feature extraction. The FIM is repeated $N$ times, and the obtained features are further handled by the convolution layers to generate the output.

In the FIM, we decouple the input feature into high- and low-frequency components. Traditional Fourier or wavelet transforms introduce notable information loss during feature extraction. Additionally, these methods exhibit high computational complexity. In this paper, we employ pooling for high- and low-frequency feature separation, which is rather computationally efficient. To illustrate the effectiveness of the FIM, we visualized the feature maps at different stages. Fig. \ref{fig_fre_compare} presents a comparative analysis of the efficacy between the FIM and conventional inter-frequency coupling methodologies.The visual comparison in Fig. \ref{fig_fre_compare} clearly demonstrates the improvements brought by the FIM. Before frequency interaction (Fig. \ref{fig_fre_compare}(b)), the feature map lacks detailed frequency information, resulting in a less discriminative representation. The conventional wavelet transform (Fig. \ref{fig_fre_compare}(c)) captures some frequency information but introduces information loss, leading to a less comprehensive feature representation. In contrast, the FIM (Fig. \ref{fig_fre_compare}(d)) effectively decouples and recouples the high- and low-frequency components, capturing the rich complementary information and leading to a more detailed and accurate feature representation.

\begin{figure}[t]
\centering
\includegraphics [width=2.5in]{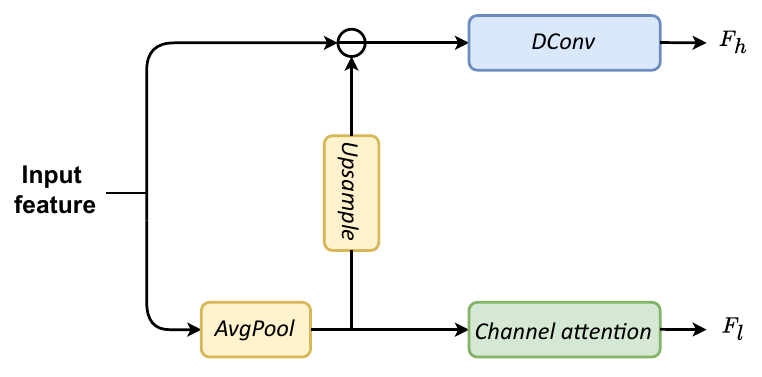}
\caption{Pooling-based frequency feature separation.}
\label{fig_pool_freq}
\end{figure}

\textbf{Pooling-Based Frequency Feature Separation.} As depicted in Fig. \ref{fig_pool_freq}, we use pooling for high- and low-frequency feature separation. Specifically, a pooling layer down-samples the input feature with dimensions $(C, H, W)$ to obtain the low-frequency feature $F_l$ at a reduced resolution $(C, \frac{H}{2}, \frac{W}{2})$. The high-frequency feature $F_h$ is computed by subtracting $F_l$ from the original features $F$. Therefore, we decouple the input features' high- and low-frequency characteristics as follows:
\begin{equation}
    F_l=\operatorname{Pooling}(F),
\end{equation}
\begin{equation}
    F_h=F-\operatorname{Upsample}(F_l),
\end{equation}
where $\operatorname{Pooling}$ denotes the down-sampling operation using average pooling with a kernel size of 2, a stride of 2, and padding of 1, and $\operatorname{Upsample}$ refers to the up-sampling operation using bilinear interpolation. After obtaining the high- and low-frequency features, we use depth-wise convolution to enhance $F_h$, and use channel attention to refine $F_l$. Through pooling-based frequency feature separation, the input HSI feature $F^H$ is decoupled into $F^H_h$ and $F^H_l$, while the input LiDAR/SAR feature $F^X$ is separated into $F^X_h$ and $F^X_l$.

\textbf{Complementary Feature Interaction.} After obtaining the high- and low-frequency features from HSI and LiDAR/SAR data, we fuse the complementary information and achieve efficient inter-frequency communication. We use a parameter-free manner to fuse the high/low-frequency features as follows:
\begin{equation}
    F^{H'}_l=F^H_l+F^X_l,
\end{equation}
\begin{equation}
    F^{X'}_h=F^H_h+F^X_h.
\end{equation}

For high- and low-frequency features within one modality, we concatenate both features and use the convolution layer to obtain the enhanced features. The output is formulated as follows:
\begin{equation}
    F^{H'}= \text{3DConv}([F^{H'}_l, F^{H}_h]),
\end{equation}
\begin{equation}
    F^{X'}= \text{Conv}([F^{X}_l, F^{X'}_h]),
\end{equation}
where 3DConv is used to exploit the spectral information of HSI. Fig. \ref{fig_frame} shows a detailed illustration of the complementary feature interaction.

\subsection{Prototype-Based Information Compensation Module}

To effectively address the challenge of multi-source feature alignment, we design the Prototype-based Information Compensation Module (PICM). We introduce two sets of learnable modality prototypes to represent the global modality information of HSI and LiDAR/SAR data, respectively. $\mathbf{P}^H$ denotes the prototype vector for HSI, while $\mathbf{P}^X$ denotes the prototype vector for SAR/LiDAR data, are initialized using a tensor of ones , and are trainable parameters throughout the network. During the model training process, the prototypes are updated via back-propagation and the Adam optimizer. Specifically, the model's loss function computes the gradient with respect to the prototypes. Subsequently, the Adam optimizer adjusts their values based on the gradient information and the specified learning rate to minimize the loss function. The prototypes play a crucial role in representing global modality information. Each prototype evolves during training to capture the essential characteristics of the corresponding modality. The training process involves updating the prototypes based on the consistency loss, which ensures that the prototypes become more representative of the global modality information. The consistency loss guides the prototypes to align with the features extracted from the input data, thereby capturing the global information of each modality. The loss function of PICNet will be elaborated in the Model Optimization section of this chapter.

The PICM is effective for multi-source feature alignment due to its ability to model global modality information and perform cross-modal feature compensation. The modality compensation fusion is achieved via cross-attention. Specifically, taking HSI feature compensation as an example, the HSI feature $\mathbf{F}^H$ obtained from frequency feature interaction encoder servers as the key $\mathbf{K}^H$ and value $\mathbf{V}^H$. Meanwhile, the SAR/LiDAR prototype vector $\mathbf{P}^X$ is utilized as the query. Formally:
\begin{equation}
\mathbf{Q}^{X} = \mathbf{P}^{X} \mathbf{W}^{Q},
\end{equation}
\begin{equation}
\mathbf{K}^{H} = \mathbf{F}^{H} \mathbf{W}^{K},
\end{equation}
\begin{equation}
\boldsymbol{V}^{H} = \mathbf{F}^{H} \mathbf{W}^{V},
\end{equation}
Here $\mathbf{W}^Q$, $\mathbf{W}^K$, and $\mathbf{W}^V$ denote the weights of linear layer. Then, the attention weights between the query and the key are calculated using the scaled dot-product attention scoring function. This computation establishes the soft correspondence between the SAR/LiDAR prototype vector $\mathbf{P}^X$ and the HSI feature  $\mathbf{F}^H$. The HSI feature $\mathbf{F}^H$ is projected into the feature space of the SAR/LiDAR data according to the attention weights, generating the compensated feature $\hat{\mathbf{F}}^X$ as follows:
\begin{equation}
\begin{aligned}
\hat{\mathbf{F}}^X & =\text{Attention}(\mathbf{Q}^X, \mathbf{K}^{H}, \mathbf{V}^{H}) \\\
& =\text{Softmax}\left(\frac{\mathbf{Q}^X \mathbf{K}^{H\top}}{\sqrt{d}}\right) \mathbf{V}^H,
\end{aligned}
\end{equation}
where $d$ is the scaling factor. Similarly, the SAR/LiDAR feature compensation works in the same way. The SAR/LiDAR feature $\mathbf{F}^X$ obtained from the frequency feature interaction encoder serves as the key $\mathbf{K}^X$ and value $\mathbf{V}^X$. The HSI prototype vector $\mathbf{P}^H$ is used as the query. Then, the compensated feature $\hat{\mathbf{F}}^H$ is computed via cross-attention.  

Next, the input HSI feature $\mathbf{F}^H$ and the compensated feature $\hat{\mathbf{F}}^X$ are combined via element-wise summation. The input SAR/LiDAR feature $\mathbf{F}^X$ and the compensated feature $\hat{\mathbf{F}}^H$ are combined as follows:
\begin{equation}
\mathbf{I}^{H}=\mathbf{F}^H+\mathbf{\hat{F}}^X,
\end{equation}
\begin{equation}
\mathbf{I}^X=\mathbf{F}^X+\mathbf{\hat{F}}^H,
\end{equation}
where $\mathbf{I}^H$ and $\mathbf{I}^X$ denote the refined HSI and SAR/LiDAR features, respectively. These feature representations encapsulate the original feature information of each source along with their aligned complementary information. This ensures effective cross-modal feature fusion and interaction.

Next, $\mathbf{I}^H$ and $\mathbf{I}^X$ are concatenated to generate the fused features as follows:
\begin{equation}
\mathbf{I}_{out}=\text{Concat}(\mathbf{I}^H, \mathbf{I}^X).
\end{equation}

Finally, $\mathbf{I}_{out}$ are fed into the refinement module. In the refinement module,  two $3\times 3$ depth-wise convolutional layers are employed to refine the details of the features, and then a Softmax layer is used to generate the classification results. By explicitly modeling and aligning cross-modal features, the PICM facilitates a more comprehensive and effective integration of multi-source remote sensing data for land cover classification.

The effectiveness of the PICM for multi-source feature alignment can be attributed to its ability to capture global modality information through learnable prototypes and perform cross-modal compensation via cross-attention. This approach not only enhances the alignment of features from different modalities but also improves the overall robustness and accuracy of the classification model.

\begin{figure*}[htbp]
\begin{center}
\includegraphics[width=0.7\linewidth]{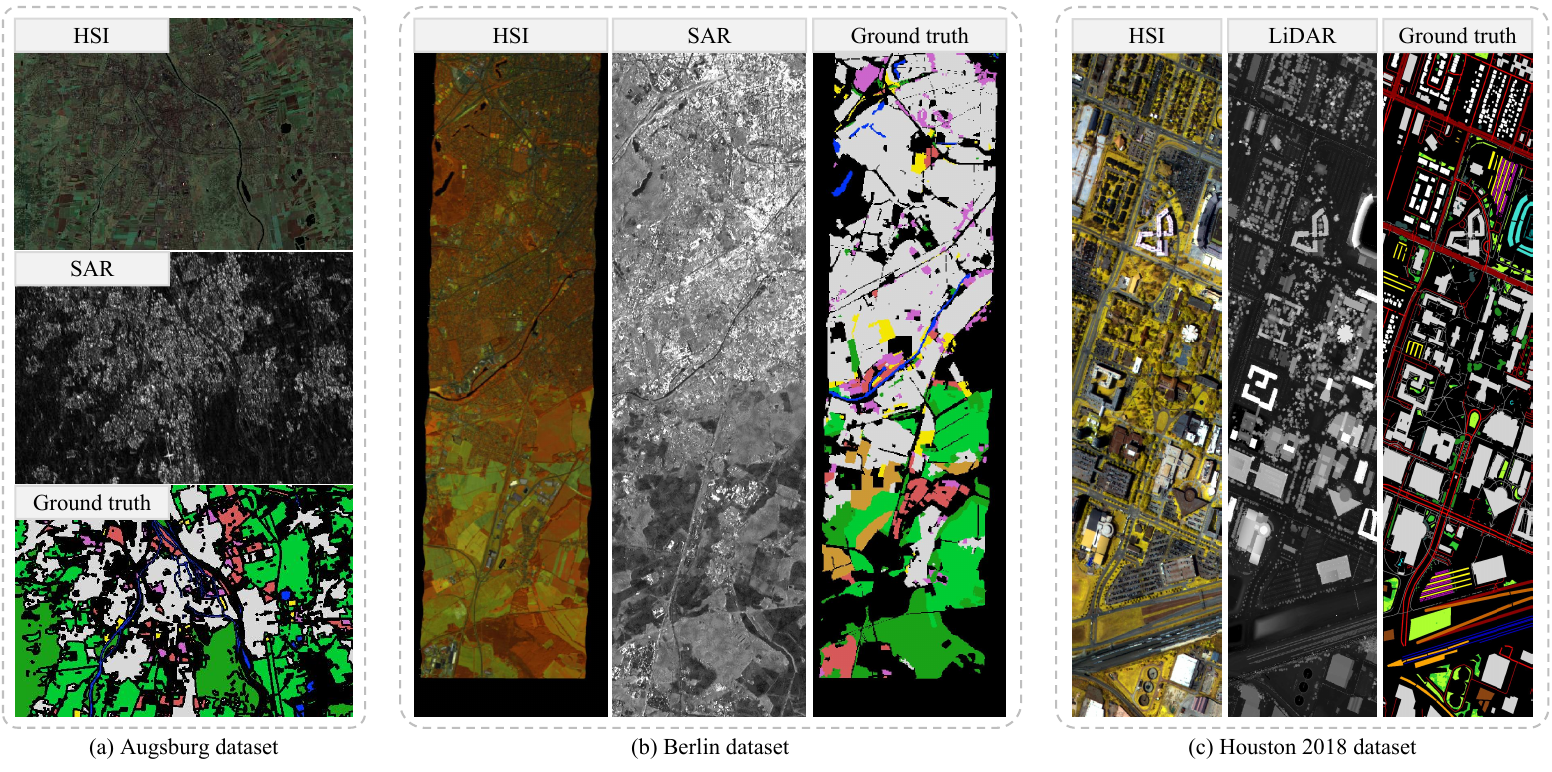} 
\end{center}
\caption{Multisource remote sensing dataset for land cover classification. (a) Augsburg dataset. (b) Berlin dataset. (c) Houston2018 dataset.}
\label{vs}
\end{figure*}

\subsection{Model Optimization}

To fully utilize the complementary aligned feature information from both modalities obtained from the prototype-based information compensation module, we design a consistency loss to guide the learning of prototype vector, which constrains the compensated feature $\hat{\mathbf{F}}^H$ and $\hat{\mathbf{F}}^H$ to be aligned with the read HSI or SAR/LiDAR features. The consistency loss $\mathcal{L}^X_{cyc}$ and $\mathcal{L}^H_{cyc}$ are defined as follows:

\begin{equation}
\mathcal{L}^X_{cyc}=\parallel\mathbf{F}^X -\hat{\mathbf{F}}^{X}\parallel_2
\end{equation}

\begin{equation}
\mathcal{L}^H_{cyc}=\parallel\mathbf{F}^H - \hat{\mathbf{F}}^H\parallel_2
\end{equation}

The consistency loss can be seen as a form of metric learning, where the goal is to minimize the distance between the compensated and original features, ensuring that the learned representations are consistent across modalities. This approach is related to contrastive learning methods, which also aim to align representations from different views or modalities by minimizing a distance metric. However, unlike contrastive learning, which often involves maximizing the distance between negative pairs, our consistency loss focuses solely on minimizing the distance between positive pairs (the original and compensated features).

By minimizing $\mathcal{L}^X_{cyc}$ and  $\mathcal{L}^H_{cyc}$, the prototype vectors are guided to effectively leverage complementary features from the corresponding modality. This enhances cross-modal feature alignment, yielding robust and reliable modality compensation. Additionally, the cross-entropy loss $\mathcal{L}_{ce}$ is employed to evaluate the classification performance by measuring the discrepancy between the classification result and the ground-truth label. 

The overall loss function $\mathcal{L}$ for the proposed PICNet combines the consistency loss and cross-entropy loss together, which is formulated as:

\begin{equation}
\mathcal{L}=\mathcal{L}_{ce}+\lambda_1\mathcal{L}^X_{cyc}+\lambda_2\mathcal{L}^H_{cyc},
\end{equation}
where $\lambda_1$ and $\lambda_2$ are the hyperparameters to balance the weight among classification loss and consistency loss. This composite loss ensures both accurate alignment of cross-modal features and precise classification performance.  

\subsection{Model Scalability}

The scalability of PICNet is crucial for its practical application in diverse remote sensing scenarios. PICNet's architecture is modular and flexible, allowing it to seamlessly integrate new data types without significant structural adjustments. The key components of PICNet, namely the Frequency Interaction Module (FIM) and the Prototype-based Information Compensation Module (PICM), are designed to be adaptable to various data types.

PICNet's input data primarily falls into two categories: spectral data and spatial data. Spectral data includes hyperspectral and multispectral data, which provide detailed spectral information about the land cover types. These data types are essential for identifying and classifying different materials based on their spectral signatures. Spatial data includes Synthetic Aperture Radar (SAR) and Light Detection and Ranging (LiDAR) data, which offer high spatial and vertical resolution. These data types are valuable for generating accurate elevation models and providing detailed spatial information about the terrain and structures. The modular design of PICNet allows it to perfectly fit these data types, achieving a plug-and-play effect.

The FIM is designed to handle different frequency components of the input data. It can be easily adjusted to accommodate new data types by modifying the pooling parameters. The PICM uses learnable modality prototypes to represent global modality information. The number of prototypes and their dimensions can be adjusted to match the characteristics of new data types. This flexibility ensures that PICM can effectively model the complementary information from diverse data sources.

\definecolor{m1}{HTML}{1AA319}  
\definecolor{m2}{HTML}{D8D8D8}  
\definecolor{m3}{HTML}{D85959}  
\definecolor{m4}{HTML}{00CC33}  
\definecolor{m5}{HTML}{CC9934}  
\definecolor{m6}{HTML}{F4E701}  
\definecolor{m7}{HTML}{CC66CC}  
\definecolor{m8}{HTML}{0035FF}  

\begin{table}[h]
\centering
\caption{Number of training and test samples in each class on the Augsbug dataset.}
\scalebox{0.85}{
\begin{tabular*}{250pt}{@{\extracolsep{\fill}}ccccc}
\hline\toprule
No. & Name & Color & Training & Test   \\
\midrule
1    & Forest & \cellcolor{m1}  & 146 & 13361       \\
2    & Residential Area & \cellcolor{m2}  & 264 & 30065       \\
3    & Industrial Area & \cellcolor{m3}  & 21 & 3830        \\
4    & Low Plants & \cellcolor{m4}  & 248 & 26609       \\
5 & Soil & \cellcolor{m5} & 331 & 17095 \\
6    & Allotment & \cellcolor{m6}  & 52 & 523       \\
7    & Commercial Area & \cellcolor{m7}  & 7 & 1638        \\
8    & Water & \cellcolor{m8}  & 23 & 1507       \\
\midrule
-    & \textbf{Total} & - & \textbf{761} & \textbf{77533}     \\
\bottomrule\hline
\end{tabular*}}
\label{3-1}
\end{table}

\begin{table}[h]
\centering
\caption{Number of training and test samples in each class on the Berlin dataset.}
\scalebox{0.85}{
\begin{tabular*}{250pt}{@{\extracolsep{\fill}}ccccc}
\hline\toprule
No. & Name & Color & Training & Test  \\
\midrule
1    & Forest & \cellcolor{m1}  & 443 & 54511       \\
2    & Residential Area & \cellcolor{m2}  & 423 & 268219       \\
3    & Industrial Area & \cellcolor{m3}  & 499 & 19067        \\
4    & Low Plants & \cellcolor{m4}  & 376 & 58906       \\
5    & Soil & \cellcolor{m5}  & 331 & 17095       \\
6    & Allotment & \cellcolor{m6}  & 280 & 13025        \\
7    & Commercial Area & \cellcolor{m7}  & 298 & 24526       \\
8    & Water & \cellcolor{m8}  & 170 & 6502       \\
\midrule
-    & \textbf{Total} & - &  \textbf{2820} & \textbf{461851}     \\
\bottomrule\hline
\end{tabular*}}
\label{3-2}
\end{table}

\definecolor{hh1}{HTML}{32CD33}  
\definecolor{hh2}{HTML}{ADFF30}  
\definecolor{hh3}{HTML}{008081}  
\definecolor{hh4}{HTML}{228B22}  
\definecolor{hh5}{HTML}{2E4F4E}  
\definecolor{hh6}{HTML}{8B4512}  
\definecolor{hh7}{HTML}{00FFFF}  
\definecolor{hh8}{HTML}{FFFFFF}  
\definecolor{hh9}{HTML}{D3D3D3}  
\definecolor{hh10}{HTML}{FE0000}  
\definecolor{hh11}{HTML}{A9A9A9}  
\definecolor{hh12}{HTML}{696969}  
\definecolor{hh13}{HTML}{8B0001}  
\definecolor{hh14}{HTML}{C86400}  
\definecolor{hh15}{HTML}{FEA500}  
\definecolor{hh16}{HTML}{FFFF00}  
\definecolor{hh17}{HTML}{DAA521}  
\definecolor{hh18}{HTML}{FF00FE}  
\definecolor{hh19}{HTML}{0000FE}  
\definecolor{hh20}{HTML}{3FE0D0}  

\begin{table}[h]
\centering
\caption{Number of training and test samples in each class on Houston 2018 dataset.}
\scalebox{0.85}{
\begin{tabular*}{250pt}{@{\extracolsep{\fill}}ccccc}
\hline\toprule
No. & Name & color & Training & Test   \\
\midrule
1    & Health grass & \cellcolor{hh1}  & 1000 & 39196       \\
2    & Stressed grass& \cellcolor{hh2}  & 1000 &  130008       \\
3    & Artificial turf& \cellcolor{hh3}  & 1000 & 2736        \\
4    & Evergreen trees& \cellcolor{hh4}  & 1000 &  54322       \\
5    & Deciduous trees& \cellcolor{hh5}  & 1000 & 20172       \\
6    & Bare earth& \cellcolor{hh6}  & 1000 &  18064        \\
7    & Water& \cellcolor{hh7}  & 500 & 1064       \\
8	 &	Residential buildings& \cellcolor{hh8}       & 1000 & 15899   \\
9	 &	 Non-residential buildings& \cellcolor{hh9}  & 1000 & 894769  \\
10	 &	Roads& \cellcolor{hh10}                       & 1000 &  183283 \\
11	 &	 Sidewalks & \cellcolor{hh11}                 & 1000 & 136035  \\
12	 &	Crosswalks & \cellcolor{hh12}                 & 1000 &  6059   \\
13	 &	Major thoroughfares& \cellcolor{hh13}         & 1000 & 185438  \\
14	 &	 Highways    & \cellcolor{hh14}               & 1000 &  39438  \\
15	 &	 Railways   & \cellcolor{hh15}                & 1000 & 27748   \\
16	 &	 Paved parking lots   & \cellcolor{hh16}      & 1000 & 45932   \\
17	 &	 Unpaved parking lots & \cellcolor{hh17}      & 250  &  587    \\
18	 &	Cars    & \cellcolor{hh18}                    & 1000 &  26289  \\
19	 &	Trains  & \cellcolor{hh19}                    & 1000 & 21479   \\
20	 &	Stadium seats  & \cellcolor{hh20}             & 1000 & 27296   \\
\midrule
-    & \textbf{Total} & - &  \textbf{18750} & \textbf{2000160}     \\
\bottomrule\hline
\end{tabular*}}
\label{3-3}
\end{table}

\section{Experimental Results and Analysis}\label{S4}

\subsection{Dataset Description and Evaluation Metrics}

\textbf{Augsburg dataset \cite{essd-15-113-2023}.}  This dataset was collected in the vicinity of Augsburg, Germany, and comprises a spaceborne HSI image and a dual-polarization polarimetric synthetic aperture radar (PolSAR) image. The HSI was acquired using the HySpex sensor, while the PolSAR data was collected by the Sentinel-1 sensor. To evaluate the performance of the proposed PICNet, we reduced the spatial resolution of all modalities to 30m GSD. The Augsburg dataset consists of 332$\times$485 pixels and 180 spectral bands, with the spectral range of the HSI image spanning from 0.4$\mu$m to 2.5$\mu$m. The SAR data includes four features: VV intensity, VH intensity, and the real and imaginary parts of the off-diagonal elements of the PolSAR covariance matrix. The ground truth data was derived from OpenStreetMap data. 
Fig. \ref{vs}(a) illustrates the distribution of various land covers in the Augsburg dataset, while the details of the training and test sets are presented in Table \ref{3-1}.

\textbf{Berlin dataset \cite{okujeni2016berlin}.}  This dataset covers some urban and rural areas of Berlin, including HSI and SAR data. The HSI is generated from HyMap airborne hyperspectral sensor. The SAR data consists of Sentinel-1 dual-Pol (VV-VH) Single-look Complex (SLC) products from the European Space Agency (ESA). The spatial resolution of the HSI is 30m GSD, consisting of 797$\times$220 pixels and 244 spectral bands. The spectrum of HSI ranges from 0.4$\mu$m to 2.5$\mu$m. The corresponding SAR data has a spatial resolution of 13.89m GSD. We use nearest neighbor interpolation to match the spatial resolution of HSI and SAR data. Fig. \ref{vs}(b) depicts the distribution of various land covers in the Berlin dataset. The details of the training and test sets for the Berlin dataset can be found in Table \ref{3-2}.

\textbf{Houston2018 dataset \cite{8328995}.}  This dataset offers comprehensive coverage of both the University of Houston campus and neighboring urban areas. It was introduced by the University of Houston during the 2018 IEEE GRSS Data Fusion Contest. The dataset contains LiDAR, HSI, and multispectral data. In this paper, we used HSI and LiDAR data for classification. The spectrum range of HSI spans from 0.38 $mu$m to 1.05 $mu$m with 48 spectral bands. Fig. \ref{vs}(c) depicts the distribution of various land covers in the Houston 2018 dataset. The details of the training and test sets for the Houston 2018 dataset can be found in Table \ref{3-3}.

\textbf{Evaluation metrics.} The evaluation metrics used in this paper include Overall Accuracy(OA), Average Accuracy(AA), and Kappa. OA is defined as the ratio of the number of correctly classified samples to the total number of samples, i.e., the ratio of the sum of the main diagonal elements of the confusion matrix to the sum of the matrix elements. OA represents the proportion of correctly classified samples out of all the samples in the dataset, providing a measure of the model's overall performance. Kappa coefficient as a consistency test index is a multivariate discrete evaluation method for the classification accuracy of remote sensing images by a comprehensive confusion matrix. Kappa not only considers the number of samples correctly classified in the main diagonal direction of the confusion matrix but also takes into account the cases of missed and misclassified samples outside the main diagonal. Therefore, Kappa reflects the overall situation and local details of the classification results more comprehensively.

\begin{figure}[h]
    \centering
    \includegraphics[width=2.5in]{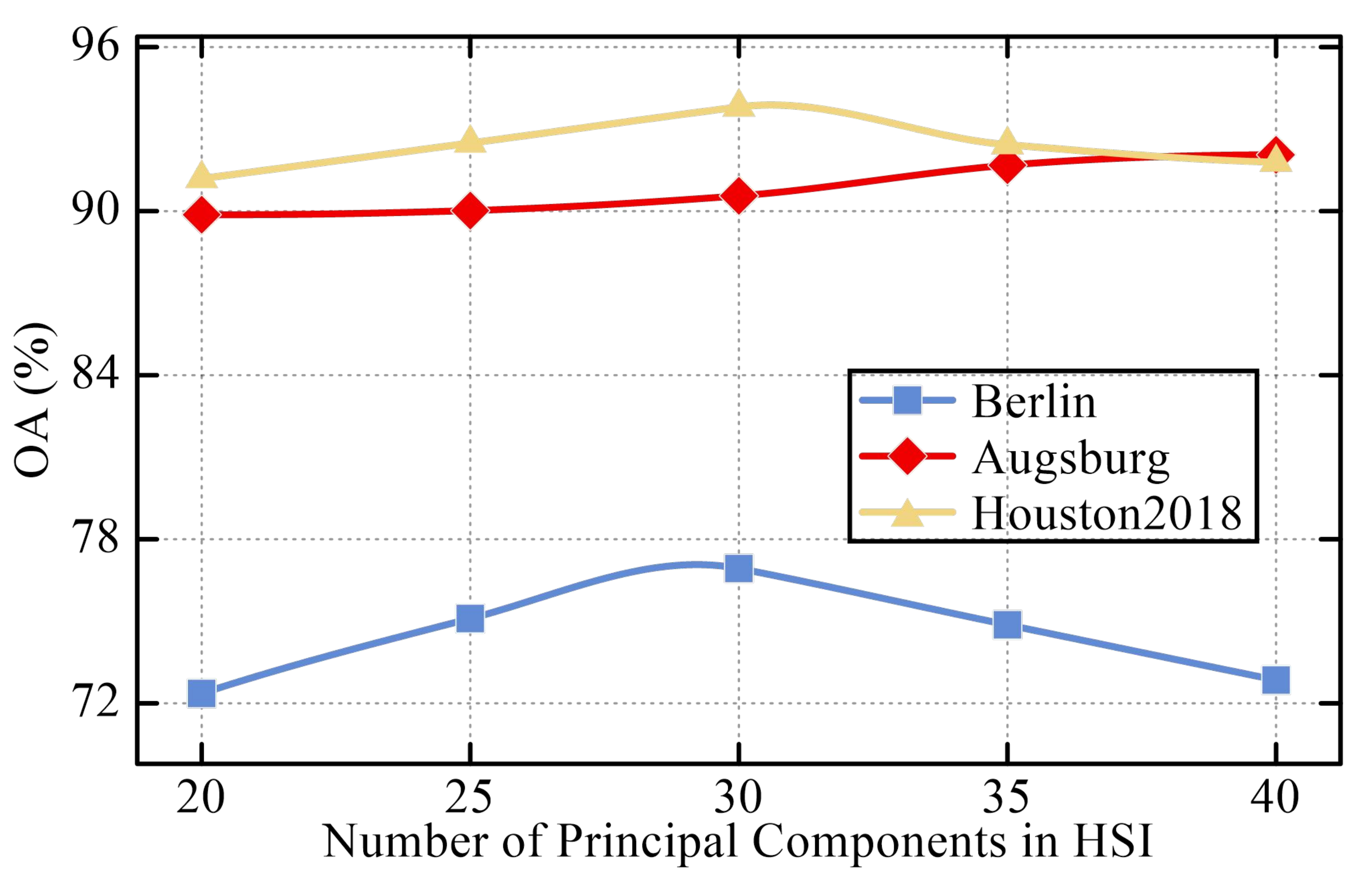}
    \caption{The relationship between OA and the number of principal components in HSI.}
    \label{fig_para_pca}
\end{figure}

\begin{figure}[h]
    \centering
    \includegraphics[width=2.5in]{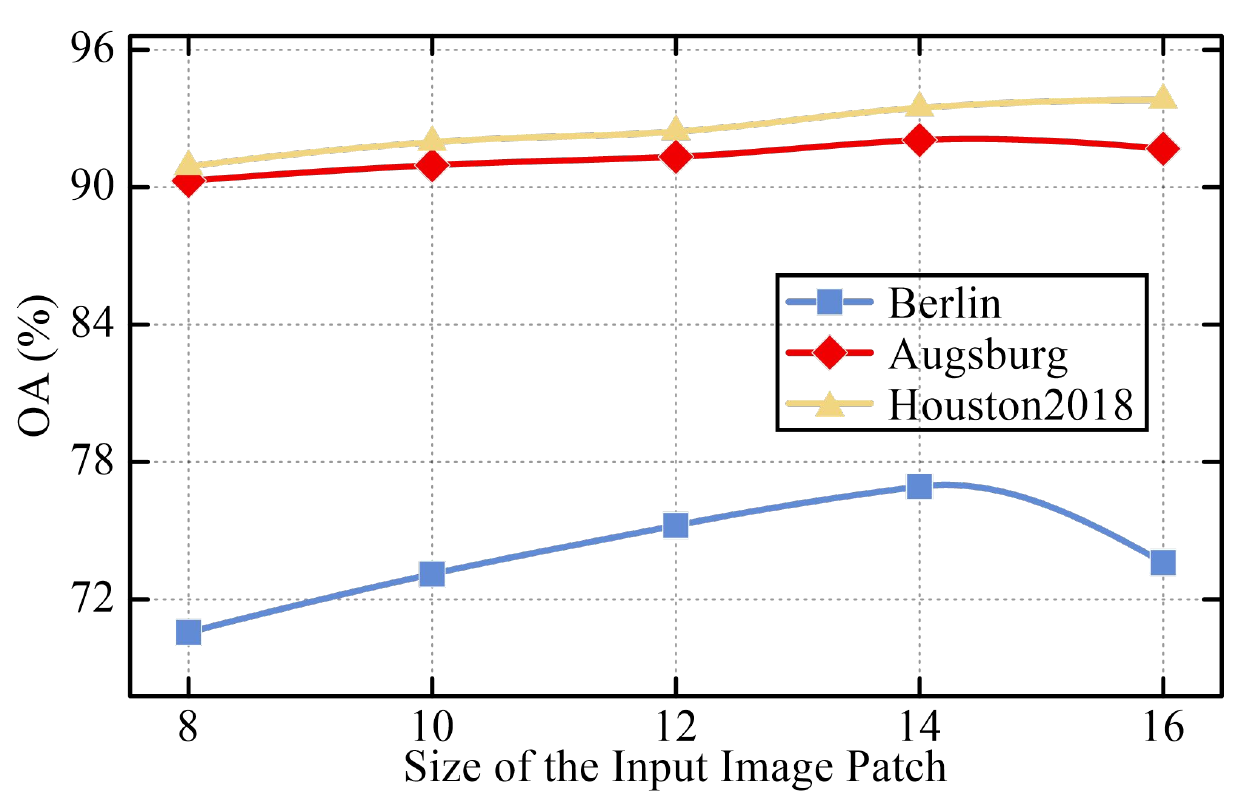}
    \caption{The relationship between OA and the size of the input image patch.}
    \label{fig_para_patch}
\end{figure}

\begin{figure}[h]
    \centering
    \includegraphics[width=2.5in]{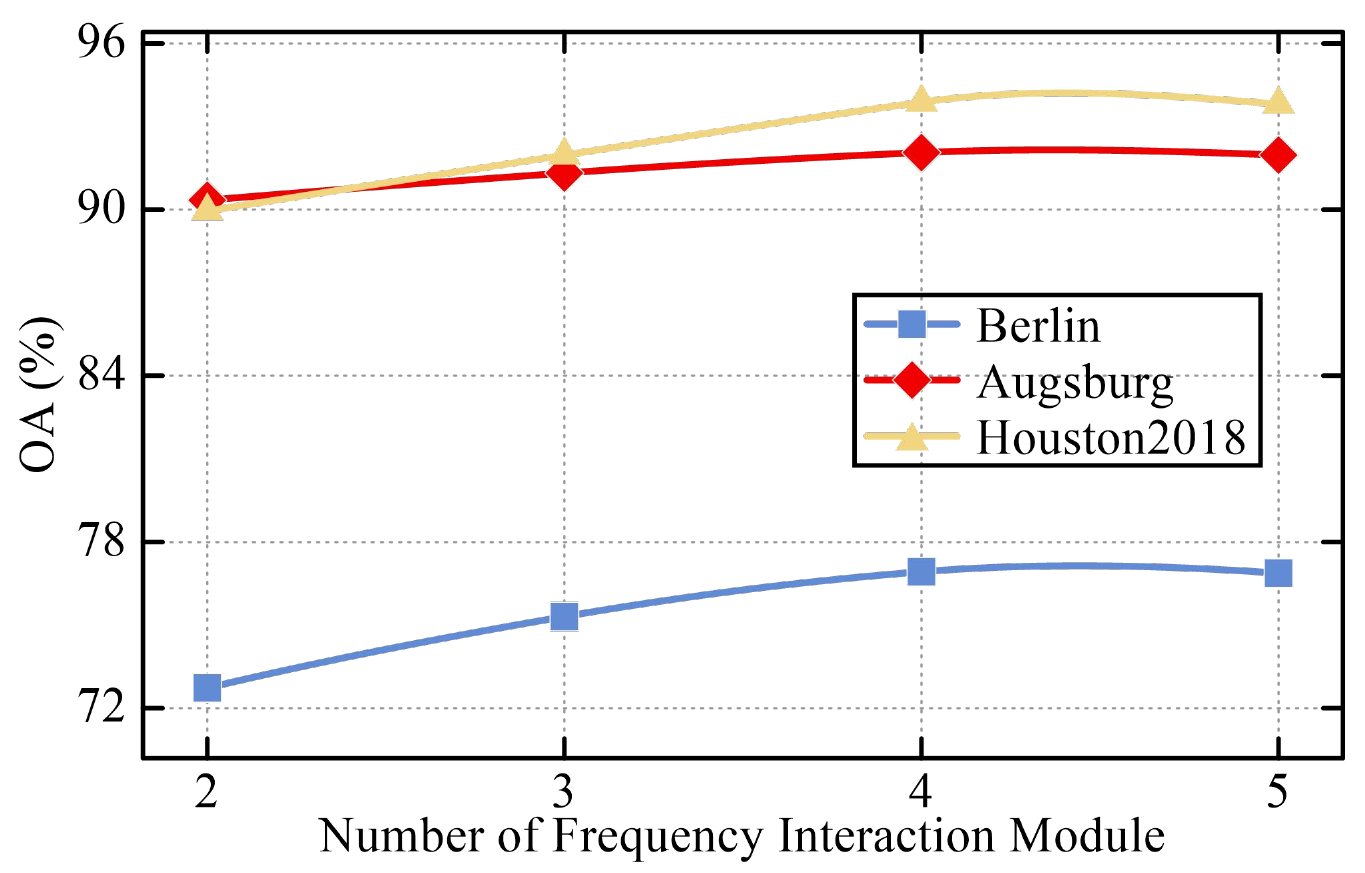}
    \caption{The relationship between OA and the number of frequency interaction modules.}
    \label{fig_para_n}
\end{figure}

\subsection{Parameter Analysis}

We provide a detailed analysis of several important parameters that may affect the classification performance of our PICNet. These parameters include the number of principal components in HSI, the size of input image patches, and the number of frequency interaction modules.

\textbf{Number of Principal Components in HSI.} The number of spectral bands of HSI in the Augsburg, Berlin, and Houston2018 datasets is 180, 244, and 48, respectively. To reduce the computational burden and the redundancy in spectral information, principal component analysis (PCA) is selected to reduce the dimensionality of the spectral dimension in HSI. The number of spectral bands $N_p$ in PCA is an important parameter that may affect the classification accuracy. We test different values of $N_p$ from 20 to 40, and the relationship between $N_p$ and OA is shown in Fig. \ref{fig_para_pca}. As can be observed that when the value of $N_p$ is 30, the proposed PICNet achieves the best performance in the Augsburg and Houston 2018 datasets. When the value of $N_p$ is set to 35, our PICNet achieves the best classification accuracy in the Berlin dataset. Therefore, for the Augsburg and Houston 2018 datasets, we use 30 principle components for HSI. For the Berlin dataset, we use 35 principle components for HSI.

\textbf{Size of the Input Image Patches.} We divided the input data into overlapping image patches and fed these patches into the network for training and testing. If the patch is too small, the network can hardly explore the contextual information. However, if the patch is too large, the classification results would be easily affected by the surrounding pixels. Hence, the size of the input image patches is a critical parameter that demands meticulous analysis. The size of input image patches $k$ was set from 8 to 16. The relationship between $k$ and OA is shown in Fig. \ref{fig_para_patch}. As can be observed on the Berlin and Houston 2018 datasets, when the value of $k$ is set to 14, our PICNet achieves the best classification results. When the value of $k$ is set to 16, our PICNet achieves the best results on the Augsburg dataset. Therefore, we can conclude that different input patch sizes affect the classification performance, and the optimal input patch size depends on the diversity of each dataset.

\begin{figure*}[ht]
\includegraphics[width=\linewidth]{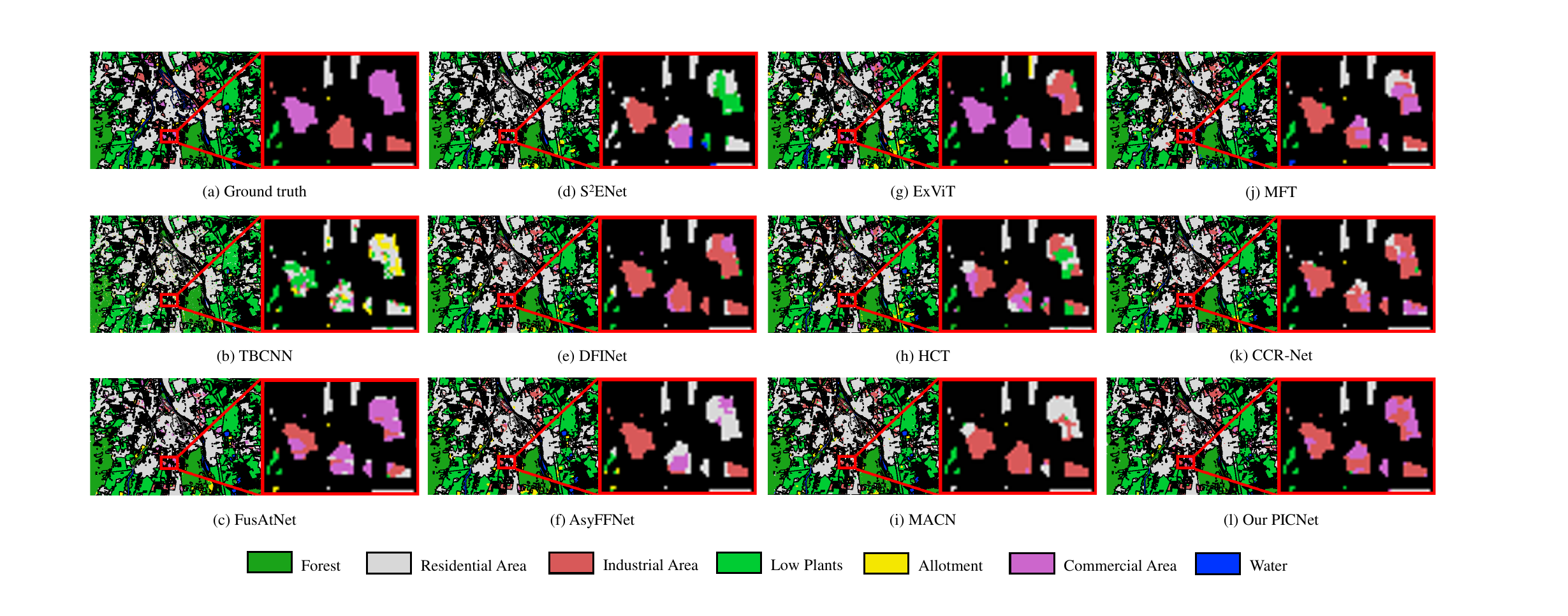} 
\caption{Classification results of different methods for the Augsburg dataset. (a) Ground truth. (b) TBCNN. (c) FusAtNet. (d) S2ENet. (e) DFINet. (f) AsyFFNet. (g) ExViT. (h) HCT. (i) MACN. (j) MFT. (k) CCR-Net. (l) Our PICNet.}
\label{fig_augsburg_vis}
\end{figure*}

\begin{table*}[t]
\centering
\caption{Classification performance of different methods on the Augsburg dataset.}
\scalebox{0.9}{
\begin{tabular}{c|cccccccccccc}
\hline\toprule
~~~Class~~~    & ~TBCNN~ & FusAtNet & ~S2ENet~ & ~DFINet~   & AsyFFNet & ExViT & HCT    & MACN & MFT & CCR-Net   & Ours\\
\midrule
Forest             & ~90.88~ & ~93.78~ & ~97.06~ & ~97.38~     & ~\textbf{97.47}~ & ~90.04~ & ~97.07~ & ~96.52~ & ~96.00~ & ~97.00~ & ~96.35~ \\ 
Residential area   & ~93.89~ & ~97.58~ & ~88.05~ & ~\textbf{98.37}~ & ~90.19~ & ~95.44~ & ~90.95~ & ~91.96~ & ~94.77~ & ~94.68~ & ~97.63~ \\ 
Industrial area    & ~8.28~  & ~26.48~ & ~70.77~ & ~61.31~   & ~56.09~ & ~34.58~ & ~60.07~ & ~60.98~ & ~53.71~ & ~34.60~ & \textbf{~71.83~} \\
Low plants         & ~91.97~ & ~97.67~ & ~96.88~ & ~92.63~   & ~\textbf{97.38}~ & ~90.68~ & ~95.57~ & ~96.64~ & ~97.03~ & ~90.64~ & ~96.55~ \\
Allotment          & ~38.24~ & ~\textbf{52.77}~ & ~41.58~ & ~49.33~ & ~21.10~ & ~51.82~ & ~15.92~ & ~23.02~ & ~42.45~ & ~45.12~ & ~26.96~ \\
Commercial area    & ~1.40~  & ~24.66~ & ~24.24~  & ~3.54~  & ~11.31~  & ~\textbf{28.63}~ & ~10.95~ & ~11.44~ & ~0.67~ & 3.62~ & ~16.36~ \\
Water              & ~10.82~ & ~47.51~ & ~81.82~ & ~26.61~ & ~\textbf{87.59}~ & ~17.65~ & ~84.64~ & ~83.92~ & ~20.37~ & ~29.83~ & ~16.36~ \\
\midrule
OA                 & ~84.53~ & ~90.62~ & ~91.14~ & ~90.66~ & ~90.44~ & ~86.65~ & ~89.91~ & ~90.83~ & ~89.94~ & ~87.21~ & ~\textbf{92.06}~ \\ 
AA                 & ~47.92~ & ~\textbf{62.92}~ & ~62.51~ & ~61.30~ & ~62.91~ & ~58.40~ & ~61.27~ & ~62.36~ & ~57.86~ & ~56.50~ & ~60.82~ \\ 
Kappa              & ~77.13~ & ~86.33~ & ~87.06~ & ~86.47~ & ~86.17~ & ~80.79~ & ~85.39~ & ~86.74~ & ~85.43~ & ~81.42~ & ~\textbf{88.51}~ \\ 
\bottomrule\hline
\end{tabular}}
\label{augsburg_compare_table}
\end{table*}

\textbf{Number of Frequency Interaction Module.} As depicted in Fig. \ref{fig_frame}, in the frequency feature interaction encoder, the Frequency Interaction Module (FIM) is repeated $N$ times for high- and low-frequency feature recoupling. The parameter $N$ is an important parameter that may affect the classification performance. We test different $N$ values on the three datasets, and the experimental results are shown in Fig. \ref{fig_para_n}. The experimental results indicate that the optimal classification results are obtained when the number of FIM is set to 4. When using more FIMs in our PICNet, the classification accuracy remains almost the same. Considering the computational burden and training efficiency, we set the number of FIMs to 4 for all three datasets in our following experiments.

\subsection{Classification Performance and Analysis}

In this subsection, several closely related methods have been implemented for performance comparison, including two-branch convolution neural network (TBCNN) \cite{tbcnn18tgrs}, FusAtNet \cite{fusatnet20cvpr}, S2ENet \cite{s2enet22tgrs}, DFINet \cite{dfinet22tgrs}, AsyFFNet \cite{asyffnet23tnnls}, ExViT \cite{exvit23tgrs}, HCT \cite{hct23tgrs}, MACN \cite{macn23tgrs}, MFT \cite{10153685} and CCR-Net \cite{9598903}. TBCNN \cite{tbcnn18tgrs} introduces a dual-branch CNN to separately extract spectral-spatial features from HSI and LiDAR data.  FusAtNet \cite{fusatnet20cvpr} leverages self-attention and cross-attention mechanisms to enhance cross-modal feature representation through the separate generation of spectral and spatial attention maps. S2ENet \cite{s2enet22tgrs} enhances the spectral representation of HSI via the spectral enhancement module, and improves the spatial representation of LiDAR data through the spatial augmentation enhancement module. DFINet \cite{dfinet22tgrs} extracts auto-correlation and cross-correlation information from multi-source feature pairs by designing a deep cross-attention module.  AsyFFNet \cite{asyffnet23tnnls} extracts cross-modal features through weight-sharing residual blocks and independent batch normalization layers, and optimizes feature fusion by incorporating sparse constraints. ExViT \cite{exvit23tgrs} classifies multi-source data through parallel position-sharing Transformer blocks and convolution blocks. It integrates a cross-modal attention module with a decision-level fusion module. HCT \cite{hct23tgrs} effectively fuses features from HSI and LiDAR data through a cross-token attention fusion encoder, leveraging the spatial context extraction capabilities of CNN and the long-range dependency modeling abilities of Transformer. MACN \cite{macn23tgrs} extracts shallow features through an adaptive CNN encoder, and integrates local and global multi-scale perception via a hybrid self-attention and convolutional Transformer. It achieves deep feature fusion by leveraging a multi-source cross-guided fusion. MFT \cite{10153685} uses multi-head cross-patch attention (mCrossPA) to fuse hyperspectral images with other multimodal data sources for improved land-cover classification. By incorporating external classification tokens from complementary data, MFT enhances feature learning and generalization, capturing long-range dependencies and boosting classification accuracy. CCR-Net \cite{9598903} uses a convolutional neural network (CNN) with a cross-channel reconstruction (CCR) module for multimodal remote sensing data classification. The CCR module enhances feature fusion by enabling information exchange across different modalities in a cross-channel manner, leading to more compact and effective feature representations.

\textit{1) Results on the Augsburg Dataset.} Table \ref{augsburg_compare_table} presents the classification performance of various methods on the Augsburg dataset. Our proposed PICNet achieves the best OA of 92.06\% and the highest Kappa coefficient of 88.51. Among the other methods, the S2ENet achieves suboptimal results since the spatial representation enhancement module improves the spatial features. TBCNN does not effectively fuse cross-modal features, and the classification accuracy is relatively low. It is evident that our PICNet effectively aligns heterogeneous multi-source features via global prototype-based information compensation. The visualized classification results of different methods are shown in Fig. \ref{fig_augsburg_vis}. After frequency feature interaction, our PICNet captures high- and low-frequency complementary features from multi-source data, producing closer results to the ground truth and thus enabling smoother classification maps.

\begin{figure*}[htbp]
\includegraphics[width=0.9\linewidth]{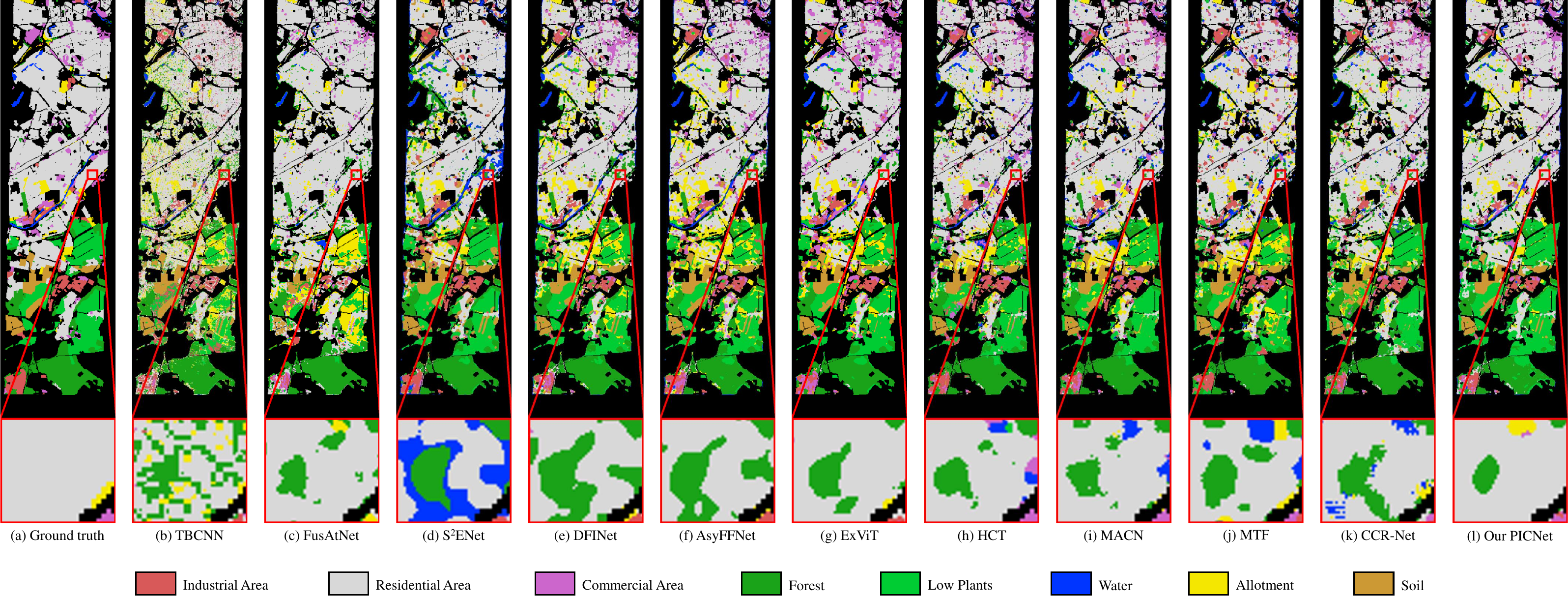} 
\caption{Classification results of different methods for the Berlin dataset. (a) Ground truth. (b) TBCNN. (c) FusAtNet. (d) S2ENet. (e) DFINet. (f) AsyFFNet. (g) ExViT. (h) HCT. (i) MACN. (j) MFT. (k) CCR-Net. (l) Our PICNet.}
\label{fig_berlin_vis}
\end{figure*}

\textit{2) Results on the Berlin Dataset.} Table \ref{berlin_compare_table} illustrates the classification performance of various methods on the Berlin dataset. Our proposed PICNet achieves improvements in OA of 9.01\%, 6.02\%, 5.67\%, 3.20\%, 3.89\%, 3.04\%, 1.42\%, 2.51\%, 4.63\% and 6.97\% compared to TBCNN, FusAtNet, S2ENet, DFINet, AsyFFNet, ExViT, HCT, MACN, MFT and CCR-Net, respectively. The detailed visualization of classification results is shown in Fig. \ref{fig_berlin_vis}. As can be observed that TBCNN and FusAtNet generate noisy classification maps. In the result of S2ENet, many regions of Residential area are falsely classified into Water. Our PICNet achieves more accurate classification results, particularly for the Low plants and Water. The training samples for these classes are relatively limited. It indicates that our PICNet method can still achieve good classification performance even when the number of training samples is limited, since the proposed PICM effectively models the global multi-source complementary information via two sets of modality-specific prototype vectors.

\begin{figure*}[htbp]
\begin{center}
\includegraphics[width=0.9\linewidth]{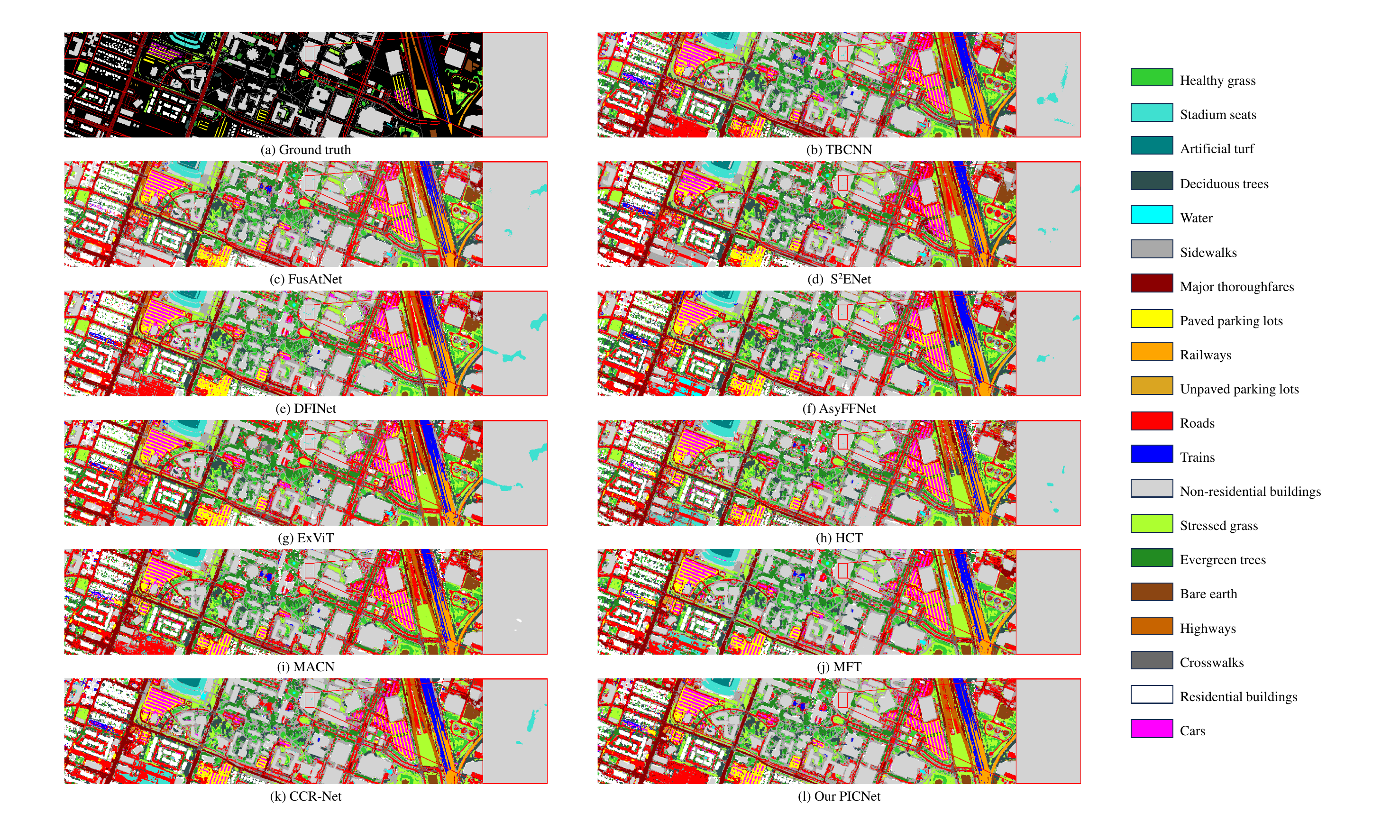} 
\end{center}
\caption{Classification results of different methods for the Houston 2018 dataset. (a) Ground truth. (b) TBCNN. (c) FusAtNet. (d) S2ENet. (e) DFINet. (f) AsyFFNet. (g) ExViT. (h) HCT. (i) MACN. (j) MFT. (k) CCR-Net. (l) Our PICNet.}
\label{fig_houston2018_vis}
\end{figure*}

\begin{table*}[htbp]
\centering
\caption{Classification performance of different methods on the Berlin dataset.}
\scalebox{0.85}{
\begin{tabular}{c|cccccccccccc}
\hline\toprule  
	~~~Class~~~      & TBCNN & FusAtNet & S2ENet & DFINet & AsyFFNet & ExViT & HCT& MACN & MFT & CCR-Net & Ours\\
 \midrule
	Forest                 & 79.51 & \textbf{86.24} & 76.22 & 76.90 & 81.02 & 62.71 & 78.56 & 77.44 & 85.84 & 84.51 & 66.24 \\ 
	Residential area       & 92.27 & 91.38 & \textbf{95.09} & 93.41 & 93.62 & 94.34 & 93.71 & 92.73 & 77.91 & 72.12 & 82.94 \\ 
    Industrial area        & 42.27 & 19.76 & 41.84 & 37.09 & 38.90 & 41.02 & 44.48 & 41.38 & \textbf{61.02} & 50.12 & 42.12 \\
    Low plants             & 72.10 & 20.00 & 68.53 & 62.76 & 71.52 & 68.79 & 72.79 & 74.56 & 62.65 & 68.81 & \textbf{90.15} \\
    Soil                   & 65.05 & 48.72 & 60.19 & 66.99 & 71.08 & 60.62 & 74.73 & 77.29 & \textbf{90.65} & 78.06 & 84.24 \\
    Allotment              & 17.59 & 38.89 & 42.56 & 46.15 & 26.04 & 26.71 & 28.21 & 24.44 & 57.28 & \textbf{65.28} & 63.26 \\
    Commercial area        & 15.49 & 18.47 & 16.42 & 18.80 & 13.23 & \textbf{35.01} & 26.28 & 25.04 & 14.75 & 30.27 & 32.04 \\
    Water                  & 66.37 & 29.61 & 68.00 & 66.94 & 66.43 & 62.14 & 43.78 & 57.49 & 46.74 & 65.50 & \textbf{78.33} \\
\midrule
    OA                     & 67.92 & 70.91 & 71.26 & 73.73 & 73.04 & 73.89 & 75.51 & 74.42 & 72.30 & 69.96 & \textbf{76.93} \\ 
    AA                     & 62.84 & 44.13 & 67.69 & 63.50 & 67.32 & 63.56 & \textbf{67.78} & 65.96 & 62.11 & 64.34 & 67.41 \\ 
    Kappa                  & 54.32 & 51.07 & 59.08 & 61.02 & 60.76 & 61.62 & 63.70 & 62.11 & 58.88 & 56.87 & \textbf{64.66} \\ 
\bottomrule\hline
\end{tabular}}
\label{berlin_compare_table}
\end{table*}

\textit{3) Results on the Houston 2018 Dataset.} The classification performance of different methods on the Houston 2018 dataset is shown in Table \ref{houston2018_compare_table}. Compared to other methods, our PICNet achieves the best performance in 14 out of 20 classes, along with an OA of 93.81\%, and Kappa coefficient of 91.99. Notably, as shown in the visualized results in Fig. \ref{fig_houston2018_vis}, our method demonstrates a clear classification advantage in the Non-residential building class. From the experiments in the Houston 2018 dataset, we conclude that in datasets with a vast amount of data, our proposed PICNet can effectively capture the high- and low-frequency information from heterogeneous multi-source data and facilitate global multi-source complementary information interaction.

\begin{table*}[htb]
\centering
\caption{Classification performance of different methods on the Houston 2018 dataset.}
\scalebox{0.85}{
\begin{tabular}{c|ccccccccccc}
\hline\toprule
~Class~                     & TBCNN & FusAtNet & S2ENet & DFINet & AsyFFNet & ExViT    & HCT   & MACN & MFT & CCR-Net   & Ours\\
\midrule
Health grass                & 87.25 & 86.20 & 71.85 & 75.46 & 94.87 & 84.55   & 74.15 & 86.24 & 89.07 & 90.91 & \textbf{97.02}\\ 
Stressed grass              & 96.62 & 93.34 & \textbf{99.25} & 95.39 & 93.89 &89.13    & 96.94 & 93.91 & 94.58 & 92.82 & 89.79\\ 
Artificial turf             & 93.99 & 93.38 & 96.55 & 97.04 & 85.01& 94.30 & 96.18 & 92.13 & \textbf{100.0} & \textbf{100.0} & \textbf{100.0}\\
Evergreen trees             & 88.95 & 92.34 & 96.10 & 90.67 & 86.89& 94.36 & 90.22 & 91.23 & 98.63 & 98.69 & \textbf{99.56}\\ 
Deciduous trees             & 67.68 & 73.69 & 95.42 & 79.39 & 69.24& 70.15 & 78.61 & 74.07 & 98.76 & 98.88 & \textbf{99.84}\\ 
Bare earth                  & 85.83 & 96.60 & 99.47 & 98.51 & 93.08& 71.79 & 95.94 & 98.68 & \textbf{100.0} & \textbf{100.0} & 99.96\\ 
Water                       & 98.26 & 99.12 & 61.57 & 93.07 & 98.43& 90.82 & 97.92 & 66.74 & \textbf{100.0} & \textbf{100.0} & \textbf{100.0}\\ 
Residential buildings       & 72.83 & 81.23 & 82.24 & 80.47 & 90.84& 80.92 & 89.35 & 84.54 & \textbf{97.75} & 95.89 & 97.08\\ 
Non-residential buildings   & \textbf{99.71} & 99.29 & 99.56 & 99.72 & 99.39& 99.68 & 99.14 & 99.42& 95.30 & 93.93 & 95.88\\ 
Roads                       & 76.15 & 85.40 & 82.39 & \textbf{89.61} & 85.87 & 88.06 & 82.86 & 85.43 & 77.49 & 73.45 & 87.58\\ 
Sidewalks                   & 73.35 & 78.06  & 81.08 & 76.19 & 79.24& 75.48 & 83.45 & 73.86 & 80.59 & 78.40 & \textbf{82.06}\\ 
Crosswalks                  & 7.70 & 22.68 & 13.56 & 22.99 & 18.02& 19.58 & 19.36 & 14.49 & \textbf{98.62} & 96.96 & 97.55\\ 
Major thoroughfares         & 84.03 & 79.88 & 77.67 & 86.89 & 84.11& 84.19 & 86.22 & 86.10 & 84.69 & 82.90 & \textbf{89.35}\\ 
Highways                    & 80.15 & 79.11 & 76.00 & 80.47 & 92.84& 80.80 & 80.09 & 80.91 & 97.64 & 98.20 & \textbf{99.23}\\ 
Railways                    & 97.34 & 96.55 & 98.75 & 98.00 & 93.70& 95.21 & 97.29 & 97.82 & 99.69 & 99.71 & \textbf{99.87}\\ 
Paved parking lots          & 90.50 & 92.07 & 96.17 & 94.03 & 85.61& 94.78 & 93.60 & 91.79 & 98.11 & 98.20 & \textbf{98.84}\\ 
Unpaved parking lots        & 90.84 & 62.29 & 78.55 & 97.68 & 47.13& 85.75 & 55.89 & 96.47 & \textbf{100.0} & \textbf{100.0} & \textbf{100.0}\\ 
Cars                        & 83.87 & 85.67 & 91.39 & 89.83 & 82.38 & 87.53 & 87.03 & 88.31 & 98.93 & 97.04 & \textbf{98.99}\\ 
Trains                      & 91.71 & 87.75 & 94.09 & 94.68 & 82.92& 94.99 & 90.53 & 96.37 & \textbf{99.95} & 99.89 & \textbf{99.95}\\ 
Stadium seats               & 91.69 & 95.76 & 95.48 & 76.02 & 98.36& 86.86 & 97.63 & 89.51 & 99.99 & 99.98 & \textbf{100.0}\\
\midrule
OA                          & 86.75 & 90.25 & 90.05 & 90.84 & 91.41& 90.17 & 91.62 & 90.42 & 92.23 & 90.70 & \textbf{93.81}\\ 
AA                          & 92.70 & 94.25 & 93.13 & 95.34 & 93.67& 94.17 & 95.10 & 94.69 & 95.49 & 94.79 & \textbf{96.63}\\ 
Kappa                       & 83.10 & 87.43 & 87.19 & 88.25 & 88.90& 87.39 & 89.15 & 87.68 & 89.97 & 88.03 & \textbf{91.99}\\ 
\bottomrule\hline
\end{tabular}}
\label{houston2018_compare_table}
\end{table*}

\subsection{Ablation Experiment}

To evaluate the effectiveness of the basic components of the proposed PICNet, we performed ablation experiments on the Augsburg, Berlin and Houston 2018 datasets. The primary objective of these experiments was to investigate the individual and combined contributions of the Frequency Interaction Module (FIM), Prototype-based Information Compensation Module (PICM), and consistency loss function to the overall performance of PICNet. Table \ref{table_ablation} shows the relationship between the different components and corresponding OA values on three datasets. We initiated the experiments with a baseline model that solely relies on convolutional neural networks (CNNs) for feature extraction and fusion. This model serves as the fundamental architecture without any of the proposed enhancements. The baseline model achieved overall accuracies (OA) of 88.51\%, 69.89\%, and 89.68\% on the Augsburg, Berlin, and Houston 2018 datasets, respectively. These results provide a reference point for evaluating the incremental improvements brought by each additional component. 

Next, we introduced the Prototype-based Information Compensation Module (PICM) to the baseline model. The PICM is designed to model the global multi-source complementary information via modality-specific prototype vectors and align the features through cross-attention computation. The results showed that the OA values improved to 90.58\%, 70.22\%, and 91.02\% on the three datasets, respectively. This indicates that the PICM effectively enhances the classification performance by addressing the challenge of inconsistent complementary information exploration during multi-source feature fusion. 

We then incorporated the Frequency Interaction Module (FIM) into the baseline model. The FIM aims to enhance the inter-frequency coupling in multi-source feature extraction by decoupling and recoupling the high- and low-frequency components. The results showed that the OA values improved to 91.36\%, 73.09\%, and 93.38\%, respectively. This demonstrates that the FIM effectively captures the inter-frequency coupling, thereby improving the overall classification performance.

Subsequently, we added the consistency loss function to the model that already includes the PICM. The consistency loss is designed to guide the learning of prototype vectors by minimizing the discrepancy between the compensated features and the original features. The results showed further improvements in the OA values to 90.99\%, 72.58\%, and 92.87\%, respectively. This highlights the effectiveness of the consistency loss in enhancing cross-modal feature alignment and improving the robustness of the model.

However, it is important to acknowledge the limitations of the PICM. While the PICM excels at aligning features through cross-attention and leveraging global complementary information, it primarily focuses on the alignment at a single frequency level. This means that the PICM may not fully capture the rich frequency characteristics present in multi-source remote sensing data, particularly the interactions between different frequency components. As a result, the PICM might struggle in scenarios where the complementary information is distributed across multiple frequency bands, leading to suboptimal feature representation and alignment.

To address this limitation, we incorporated the Frequency Interaction Module (FIM) into the model. We combined the FIM and PICM in the model. This configuration aims to leverage both the inter-frequency coupling and the global multi-source complementary information. The results showed that the OA values reached 91.87\%, 75.87\%, and 93.43\%, respectively. This indicates that the combined effect of FIM and PICM significantly improves classification performance by addressing both inter-frequency coupling and multi-source feature alignment. The FIM enhances the inter-frequency coupling in multi-source feature extraction by decoupling the features into high- and low-frequency components and then recoupling them. This process allows the model to capture the interactions between different frequency components more effectively, thereby improving the overall feature representation. By integrating the FIM with the PICM, we can leverage the strengths of both modules: the PICM for global multi-source alignment and the FIM for capturing inter-frequency interactions. The combined approach ensures more comprehensive and robust feature extraction and alignment, leading to further improvements in classification performance.

Finally, we integrated all three components—FIM, PICM, and consistency loss—into the full PICNet model. The results showed that the OA values reached the highest levels of 92.06\%, 76.93\%, and 93.81\% on the three datasets, respectively. This demonstrates that the combined contributions of FIM, PICM, and consistency loss are essential for achieving optimal performance in multi-source remote sensing data classification.

\begin{table}[]
\centering
\caption{Influence of FIM, PICM, and loss function on classification results of our PICNet}
\scalebox{0.85}{
\begin{tabular}{ccc|ccc} 
\hline\toprule
FIM &  PICM & loss & Augsburg & Berlin & Houston2018 \\
\midrule
$\times$ & $\times$ & $\times$ & 88.51 & 69.89 & 89.68\\ 
$\times$ & \checkmark & $\times$ & 90.58 & 70.22 & 91.02\\ 
\checkmark & $\times$ & $\times$ & 91.36 & 73.09 & 93.38\\ 
$\times$ & \checkmark & \checkmark & 90.99 & 72.58 & 92.87  \\ 
\checkmark & \checkmark & $\times$ & 91.87 & 75.87 & 93.43\\ 
\checkmark & \checkmark & \checkmark & \textbf{92.06} & \textbf{76.93} & \textbf{93.81} \\   
\bottomrule\hline
\end{tabular}}
\label{table_ablation}
\end{table}

\begin{figure}
    \centering
    \includegraphics[width=0.7\linewidth]{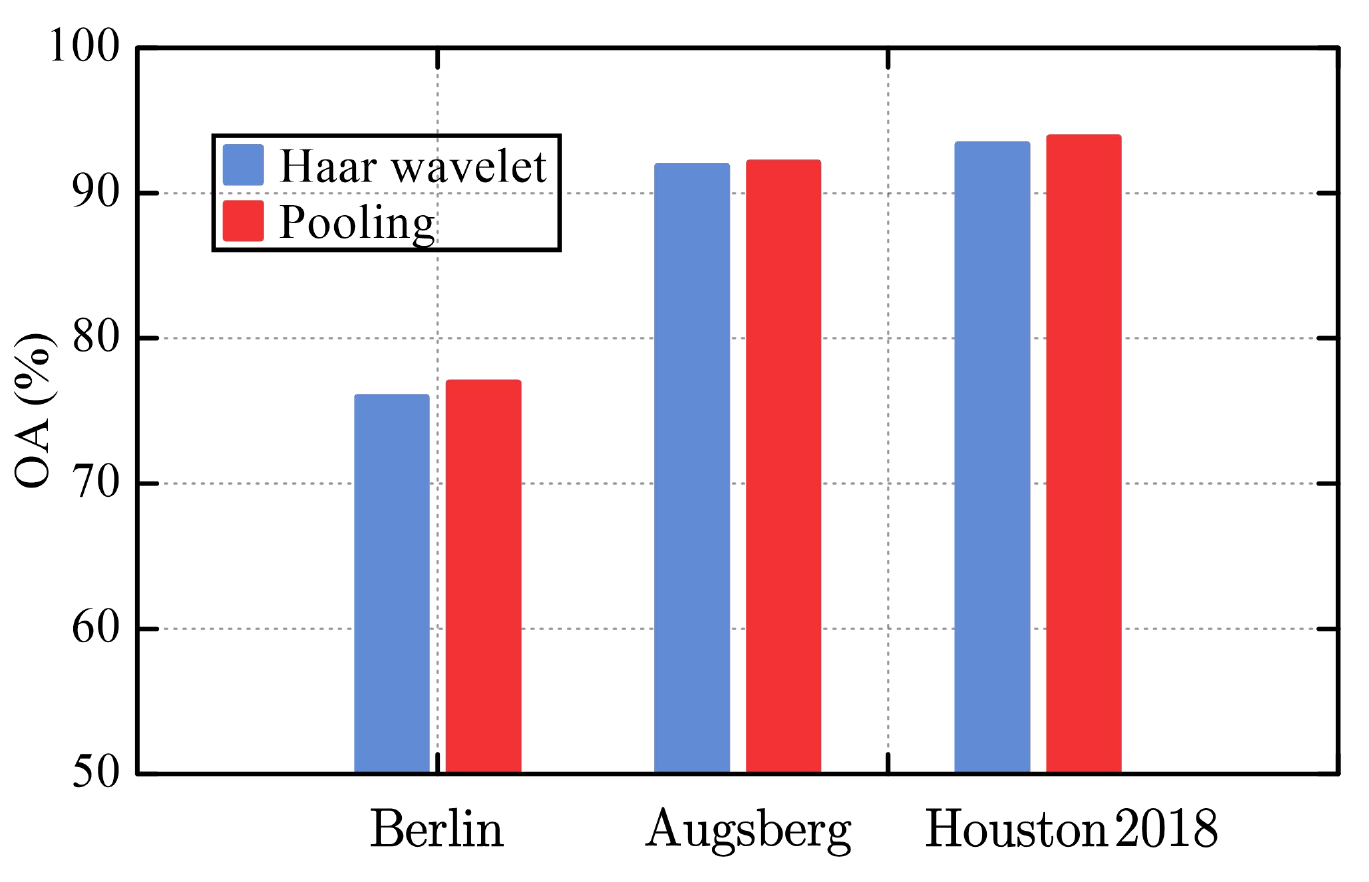}
    \caption{Influence of frequency feature separation on classification results of our PICNet.}
    \label{fig_ablation_freq}
\end{figure}

To further validate the effectiveness of the high- and low-frequency feature separation, we conducted a comparative experiment with the Haar wavelet. As shown in Fig. \ref{fig_ablation_freq}, we selected the Haar wavelet as a representative frequency-domain feature transformation method for comparison. The blue bar denotes the PICNet with Haar wavelet for frequency feature separation, while the red bar denotes the PICNet with pooling-based frequency separation. The experimental results show that the proposed method slightly improves the classification performance by leveraging the pooling-based frequency feature separation. The pooling-based method effectively reduced computational complexity, thereby achieving superior classification performance. 

\subsection{Analysis of Computational Complexity}

To provide a comprehensive computational complexity analysis, we evaluated the model parameters, floating point operations (FLOPs), and inference time of various methods on the Augsburg dataset, as shown in Table \ref{table_complex}. Our proposed PICNet model demonstrates a good balance in terms of model parameters, computational load, and inference time.

The model parameters of PICNet amount to 2.4307 M, which is at a moderate level. It is neither too large to cause resource waste nor too small to affect the model performance. In terms of FLOPs, the computational load of PICNet is 0.1291 G, also within a reasonable range. This indicates that PICNet has good control over computational resource consumption and will not slow down system efficiency due to complex calculations. As for inference time, PICNet takes 0.4771 seconds, showing a clear advantage over some models. This means that in practical applications, PICNet can provide results more quickly and enhance user experience.

The efficient use of computational resources and the good balance with performance in PICNet highlight its potential and applicability for practical deployment. It can meet the demands of complex tasks for model performance while also considering the rational use of computational resources and the need for quick responses. This lays a solid foundation for its subsequent practical application and promotion.

\begin{table*}[h]
\centering
\caption{Comparative Analysis of Model Parameters, FLOPs and Inference time on the Augsburg Dataset}
\scalebox{0.9}{
\begin{tabular}{c|cccccccccccc}
\hline\toprule
Metrics  & FusAtNet & $S^2$ENet & DFINet & AsyFFNet & ExVit  & HCT    & MACN & MFT & CCR-Net & PICNet &  \\ 
\midrule
Params (M) & 37.7177  & 1.4701    & 1.3155 & 2.4132   & 1.8848 & 4.7811 & 2.0199 & 0.2231 & 0.0222 & 2.4307   &  \\
FLOPs (G)  & 3.5619   & 0.1779    & 0.1191 & 0.2064   & 0.2901 & 0.0437 & 0.1672 & 0.0026 & 0.0003 & 0.1291  &  \\
Inference time (s) & 0.2469  & 0.2489  & 0.3182 & 0.2924 & 0.3271 & 0.7524 & 0.1917 & 0.7526 & 0.1791 & 0.4771  &\\
\bottomrule\hline
\end{tabular}}
\label{table_complex}
\end{table*}

\section{Conclusions and Future Work}

In this paper, we propose PICNet for land cover classification based on HSI and SAR/LiDAR data. Initially, we found two challenges in existing works: limitations in inter-frequency coupling and inconsistencies in complementary information exploration. To address these issues, our PICNet includes two main strategies: (1) Frequency feature interaction encoder is designed for multi-source feature extraction, which leverages FIM to enhance the high- and low-frequency feature coupling. Different frequency components from HSI and SAR/LiDAR data are recoupled for seamless inter-frequency information communication. (2) PICM is developed for global multi-source complementary information modeling. It uses two sets of modality-specific prototype vectors for multi-source feature alignment. Cross-attention computation is conducted between the prototype vectors and raw feature representations. Experimental results on three benchmark datasets demonstrate the effectiveness of our proposed PICNet.

In the future, we will focus on several promising directions to further enhance the classification performance of PICNet. First, we plan to incorporate multi-scale frequency information to improve the model's robustness in complex and noisy scenarios. This will allow the model to capture a broader range of frequency components, leading to more comprehensive feature representation. Second, we aim to enhance the cross-attention mechanism by integrating hierarchical attention structures. This will enable the model to better capture both global and local dependencies across modalities, improving feature alignment and classification accuracy. Additionally, we will explore advanced fusion techniques, such as deep fusion networks and Mamba-based architectures, to more effectively integrate multi-source data. Finally, we will address challenges like data imbalance and limited training data through techniques like data augmentation and class balancing. Evaluating PICNet on larger and more diverse datasets, such as hyperspectral and infrared data, will also be a priority to assess its generalizability and identify further areas for improvement.

\bibliography{ref}
\bibliographystyle{IEEEtran}

\end{document}